\documentclass[aps,pra,twocolumn,superscriptaddress]{revtex4-1}
\usepackage[utf8]{inputenc}
\usepackage{soul}
\usepackage{textcomp}
\usepackage[english]{babel}
\usepackage{graphicx}
\usepackage{dcolumn}
\usepackage{bm}
\usepackage{amsmath}
\usepackage{verbatim}     
\usepackage[dvipsnames]{xcolor}       
\usepackage[normalem]{ulem}   
\usepackage{bbold}
\usepackage{mathrsfs}
\usepackage[mathscr]{eucal}
\usepackage{physics}
\usepackage{algorithmic}

\usepackage{hyperref}
\hypersetup{
   pdfpagemode=None,
   pdfstartpage=1,
   pdfmenubar=true,
   pdftoolbar=true,
   colorlinks = true,
   linkcolor=blue,
   citecolor=blue,
   urlcolor=blue,
   bookmarksopen=false
 } 

 \usepackage{booktabs}
 \usepackage{float}

\begin{document}

\title{Quantum Process Tomography of Structured Optical Gates \\with Convolutional Neural Networks}

\author{Tareq Jaouni}
\affiliation{Nexus for Quantum Technologies, University of Ottawa, K1N 5N6, Ottawa, ON, Canada}

\author{Francesco Di Colandrea}
\email{francesco.dicolandrea@uottawa.ca}
\affiliation{Nexus for Quantum Technologies, University of Ottawa, K1N 5N6, Ottawa, ON, Canada}

\author{Lorenzo Amato}
\affiliation{Condensed Matter Theory Group, Paul Scherrer Institute, CH-5232 Villigen PSI, Switzerland}
\affiliation{Laboratory for Solid State Physics, ETH Zurich, CH-8093 Zurich, Switzerland}

\author{Filippo Cardano}
\affiliation{Dipartimento di Fisica, Universit\`{a} di Napoli Federico II, Complesso Universitario di Monte Sant'Angelo, Via Cintia, 80126 Napoli, Italy}

\author{Ebrahim Karimi}
\affiliation{Nexus for Quantum Technologies, University of Ottawa, K1N 5N6, Ottawa, ON, Canada}
\affiliation{National Research Council of Canada, 100 Sussex Drive, Ottawa ON Canada, K1A 0R6}


\begin{abstract} 
The characterization of a unitary gate is experimentally accomplished via Quantum Process Tomography, which combines the outcomes of different projective measurements to reconstruct the underlying operator. The process matrix is typically extracted from maximum-likelihood estimation. Recently, optimization strategies based on evolutionary and machine-learning techniques have been proposed. Here, we investigate a deep-learning approach that allows for fast and accurate reconstructions of space-dependent SU(2) operators, only processing a minimal set of measurements. We train a convolutional neural network based on a scalable U-Net architecture to process entire experimental images in parallel. 
Synthetic processes are reconstructed with average fidelity above 90\%. The performance of our routine is experimentally validated on complex polarization transformations. Our approach further expands the toolbox of data-driven approaches to Quantum Process Tomography and shows promise in the real-time characterization of complex optical gates.
\end{abstract}

\maketitle


\section{Introduction}
Quantum Process Tomography (QPT) is the extension of system identification to the quantum realm~\cite{Chuang1997}. A quantum operator can be characterized by measuring how a set of inputs evolve under its action, the same way a dynamical system is classified based on measured outputs~\cite{stoica}. 
For experimental purposes, the tomography is crucial in verifying the proper functionality of a quantum device. It found applications in various experiments, from nuclear magnetic resonances~\cite{PhysRevA.64.012314} to cold atoms~\cite{PhysRevA.72.013615}, trapped ions~\cite{PhysRevLett.92.220402,PhysRevLett.97.220407},  and photonic setups~\cite{PhysRevLett.91.120402,Altepeter2003,PhysRevLett.93.080502,Lobino2008, Bongioanni2010,Rahimi-Keshari2013,Zhou:15,Anton2017,PhysRevA.98.052327,Bouchard2019,DiColandreaQPT,goel2024inverse}. 
As in the case of state tomography, some optimization routine is typically needed to solve the problem of non-physical reconstructions, which are a consequence of experimental noise~\cite{James2001}. 

In photonic setups, the characterization of a set of waveplates acting on light polarization is a practical challenge. This scenario can be one-to-one mapped into the reconstruction of an SU(2) gate acting on a single qubit, which allows formulating the optical problem within the same paradigm of QPT~\cite{Aiello2006}.

In this paper, we address the more challenging scenario of characterizing optical SU(2) gates that are space-dependent. These operations are relevant in all applications requiring a local control of light polarization, such as polarization imaging~\cite{Solomon1981}, multiplexing~\cite{Davis2005}, as well as the generation and manipulation of vector beams~\cite{Zhan:09, Rosales-Guzmán_2018}. Assuming photons propagate along $z$, QPT is required at each transverse position $(x,y)$, hereafter referred to as \emph{pixel}. Iterative ${\text{pixel-by-pixel}}$ solutions prove to be not optimal, as the computation time grows linearly with the number of pixels. Moreover, similar approaches overlook the overall pixel distributions, preventing the possibility of extracting richer information from the entire experimental images. 

In order to tackle the increasing complexity, optimization strategies based on evolutionary methods and supervised learning have been recently illustrated in Ref.~\cite{DiColandreaQPT}, where an optimal set of five polarimetric measurements has also been demonstrated. Here, instead, we adopt a convolutional neural network (CNN) to deliver real-time space-resolved reconstructions of complex unitary operators. The network is trained to associate a minimal set of polarimetric images with the process parameters.

First, we test our scheme on synthetic experiments. Then, we compare its performance with the genetic routine devised in Ref.~\cite{DiColandreaQPT}, taking into account both the reconstruction's timing and accuracy. Finally, our architecture is validated experimentally on different combinations of liquid-crystal metasurfaces (LCMSs)~\cite{Rubano2019,DiColandrea2023}, realizing space-dependent polarization transformations. This ultimately demonstrates the robustness of our approach to real experiment noise.
\section{Theory}
\label{sec:theory}
A qubit rotation of an angle 2$\Theta$ around the axis ${\bm{n}=(n_x,n_y,n_z)}$, with ${0\leq \Theta<\pi}$ and ${\abs{\bm{n}}=1}$, is described by an SU(2) operator
\begin{equation}\label{eqn:SU2}
    \hat{U}=e^{-i\Theta\bm{n}\cdot\boldsymbol{\sigma}}=\cos(\Theta)\sigma_0-i\sin(\Theta)(\bm{n}\cdot\bm{\sigma}),
\end{equation}
where $\sigma_0$ is the 2$\times$2 identity matrix and ${\bm{\sigma}=(\sigma_x,\sigma_y, \sigma_z)}$ is the vector of the three Pauli matrices. 
The gate tomography is typically performed by processing an overcomplete set of projective measurements of the form
\begin{equation}\label{eqn:polmeas}
    I_{ab}=\abs{\mel{b}{\hat{U}}{a}}^2,
\end{equation}
where $\ket{a}$ and $\ket{b}$ are extracted from the three sets of states forming Mutually Unbiased Bases (MUBs) of SU(2)~\cite{Chuang1997}. The process matrix is then retrieved via a maximum-likelihood approach, i.e., by minimizing a cost function expressing the distance between the experimental outcomes $I_{ab}^{\text{exp}}$ and the corresponding theoretical predictions $I_{ab}^{\text{th}}$ \cite{PhysRevLett.93.080502,Aiello2006}:
\begin{equation}
    \mathcal{L}=\sum_{ab} (I_{ab}^{\text{exp}}-I_{ab}^{\text{th}})^2.   
\end{equation}
%
This routine is inefficient when it is independently executed on multiple gates depending on some external parameters, such as space-dependent processes.

Specifically, a space-dependent process can be modeled as a functional $\hat{U}(x,y)$ mapping the transverse plane to the SU(2) group: 
\begin{equation}
    \hat{\mathcal{U}}=\sum_{(x,y)}\hat{U}(x,y)\dyad{x,y},
    \label{eqn:pixeldecomposition}
\end{equation}
where we have decomposed the complex unitary operator $\hat{\mathcal{U}}$ in local SU(2) operators. 

The polarization of photons provides a natural way of encoding qubits, which can be manipulated via optical waveplates. In the circular polarization basis, where ${\ket{L}=(1,0)^T}$ and ${\ket{R}=(0,1)^T}$ are left and right circular polarization states, respectively, a waveplate $R_{\delta,\theta}$ having birefringence $\delta$ and optic axis oriented at an angle $\alpha$ with respect to the horizontal direction can be expressed in the matrix form
\begin{equation}
    R_{\delta,\alpha}=\begin{pmatrix}\cos{(\delta/2)} & i\sin{(\delta/2)}e^{-2i \alpha}\\i\sin{(\delta/2)}e^{2i \alpha} & \cos{(\delta/2)}\end{pmatrix}.\label{eqn:su2matrix}
\end{equation}
Here, $T$ stands for the transpose operator. A single waveplate thus implements a qubit rotation of an angle $-\delta/2$ around the equatorial axis ${\bm{n}=(\cos{2\alpha},\sin{2\alpha},0)}$ (cf.~Eq.~\eqref{eqn:SU2}). Nevertheless, more general operations can be realized by cascading multiple waveplates~\cite{SIMON1990165,Sit_2017}. Accordingly, characterizing a set of waveplates is mathematically equivalent to performing the QPT of an SU(2) operator. In this case, the measurements of Eq.~\eqref{eqn:polmeas} are realized as polarimetric measurements, involving the Stokes states $\ket{L}$ and $\ket{R}$, ${\ket{H}=\left(\ket{L}+\ket{R}\right)/\sqrt{2}}$ and ${\ket{V}=\left(\ket{L}-\ket{R}\right)/\sqrt{2}}$ (horizontal and vertical polarizations, respectively), ${\ket{D}=\left(\ket{L}+i\ket{R}\right)/\sqrt{2}}$ and ${\ket{A}=\left(\ket{L}-i\ket{R}\right)/\sqrt{2}}$ (diagonal and antidiagonal polarizations, respectively). Space-dependent processes can be realized via waveplates exhibiting a patterned optic-axis orientation, ${\alpha=\alpha(x,y)}$. 
\begin{figure*}[t!]
    \centering
    \includegraphics[width=0.8\linewidth]{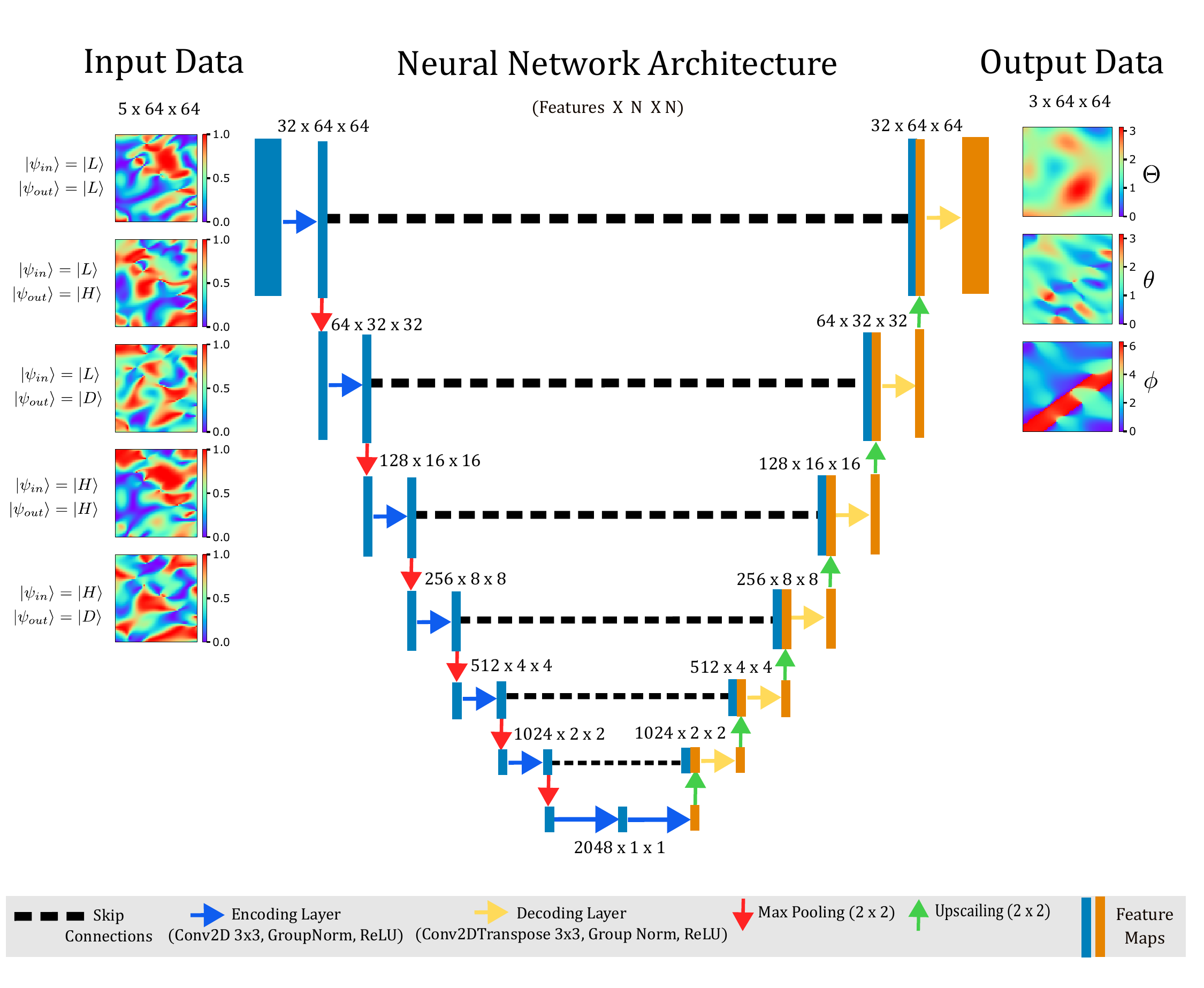}
    \caption{\textbf{Neural network architecture.} A fully convolutional neural network based on the auto-encoder U-Net is employed. Our input consists of five polarimetric measurements with the target resolution. The data is processed through a series of compression steps, each implemented as a ${2 \times 2}$ Max Pooling layer. After the bottommost layer, the data is decompressed with ${2 \times 2}$ upsampling layers. 
    The output of each encoding layer is passed as input to the decoding layer within the same level of the architecture via skip connections, which help circumvent vanishing gradients during training. The feature maps are colored blue and orange to indicate, respectively, the encoding and decoding blocks of the U-Net. 
    }
    \label{fig:fig_nnconcept}
\end{figure*}
The tomography thus consists of determining the process parameters from a set of polarimetric measurements. In our synthetic and experimental realizations, a minimal set of five measurements is considered: ${\lbrace{I_{LL}, I_{LH},I_{LD},I_{HH},I_{HD}\rbrace}}$~\cite{DiColandreaQPT}. To the best of our knowledge, this is the first implementation of an optimal QPT of SU(2) gates.
\begin{figure*}[!ht]
    \centering\includegraphics[width=0.9\linewidth]{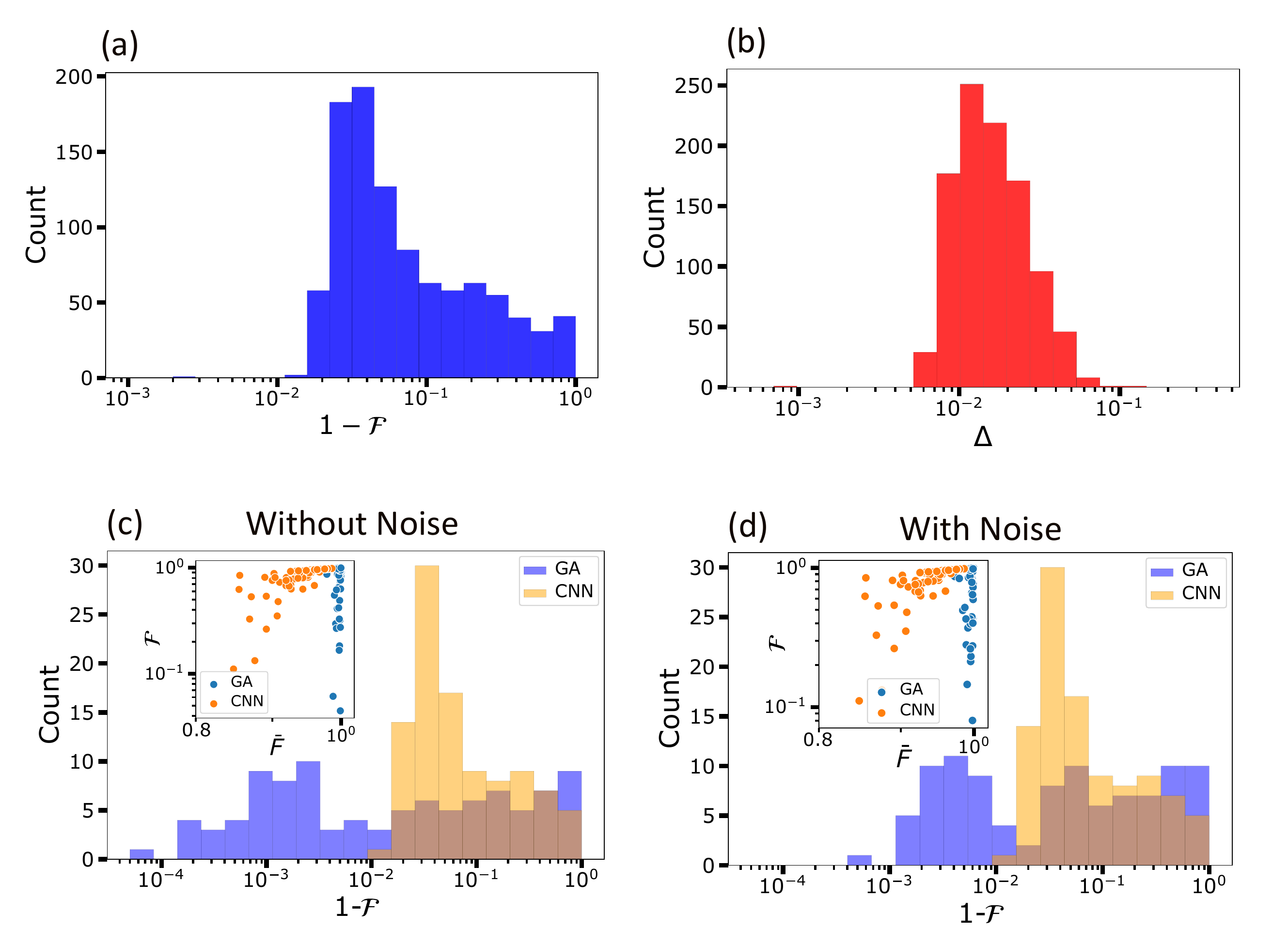}\caption{\textbf{Synthetic experiments.} Distribution of (a) the map infidelity ${1-\mathcal{F}}$ and (b) the polarimetric infidelity $\Delta$, resulting from the network reconstruction of $10^3$ complex synthetic processes. The performance of the network is compared with the genetic algorithm of Ref.~\cite{DiColandreaQPT} on a subset of 100 processes, in the case of (c) ideal measurements and (d) noisy input data ${\left(\sigma=0.02\right)}$. The insets show the correlation between pixel and map fidelity. Limited by a pixel-by-pixel approach, the genetic algorithm cannot capture global features, which leads to no correlations.}
    \label{fig:figure_of_merit}
\end{figure*}

As discussed in previous studies~\cite{Flaschner2016,Tarnowski2019,DiColandreaQPT,PhysRevResearch.5.L032016}, the main limitation of non-interferometric approaches is that these cannot capture any global phase. For this reason, we cannot distinguish a process $\hat{U}$, having parameters $(\Theta,\bm{n})$, from ${e^{i\pi}\hat{U}=-\hat{U}}$, having parameters ${(\pi-\Theta,-\bm{n})}$, as these generate the same experimental outcomes. We discuss how to solve this ambiguity in the next section. 
\section{Methodology}
\subsection{Parametrization of the training samples}
\label{main:parameterization}
Our network processes a set of five polarimetric images having resolution ${N\times N}$, where $N=64$ pixels. For each parameter describing an SU(2) process, we generate a random continuous two-dimensional function $f(x,y)$ via a discrete Fourier decomposition: 
\begin{equation}
\begin{split}
f(x,y)=\sum_{i=0}^{\Omega_x}\sum_{j=0}^{\Omega_y}c^{(1)}_{ij}\cos{\dfrac{2\pi i x}{N}}\cos{\dfrac{2\pi j y}{N}}+&\\c^{(2)}_{ij}\cos{\dfrac{2\pi i x}{N}}\sin{\dfrac{2\pi j y}{N}}+&\\c^{(3)}_{ij}\sin{\dfrac{2\pi i x}{N}}\cos{\dfrac{2\pi j y}{N}}+&\\c^{(4)}_{ij}\sin{\dfrac{2\pi i x}{N}}\sin{\dfrac{2\pi j y}{N}},
\label{eqn:parametrizationsamples}
\end{split}
\end{equation}
where all the coefficients $c^{(m)}_{ij}$
are uniformly extracted from the range ${[-1,1]}$. The maximal frequencies $\Omega_{x}$ and $\Omega_{y}$, respectively for the $x$ and $y$ axes, are independently sampled from the range $[0,5]$, thereby ensuring high-resolution pixelation: ${\max{(\Omega_{x,y})}\ll N}$.
The function is then rescaled to the ranges ${[0,\pi]}$ and ${[-1,1]}$, respectively for the rotation angle $\Theta$ and the components of the vector $\bm{n}$. Finally, the vector $\bm{n}=(n_x,n_y,n_z)$ is normalized at each transverse position. 


We also include a possible rotation of the reference frame:
\begin{equation}
\begin{split}
x &\rightarrow x'=\cos{\xi}\,x-\sin{\xi}\,y,\\
y &\rightarrow y'=\sin{\xi}\,x+\cos{\xi}\,y,
\end{split}
\end{equation}
where $\xi$ is an angle extracted from a uniform distribution in the range $[-\xi_{M},\xi_{M}]$, with $\xi_{M}=5^\circ$.
To remove the ambiguity between processes $U$ and $-U$, if the first pixel of the $n_z$ component turns out to be negative, 
the following rule is applied:
\begin{equation}
\begin{split}
\Theta(x,y)&\rightarrow \pi-\Theta(x,y),\\
n_x(x,y)&\rightarrow -n_x(x,y),\\
n_y(x,y)&\rightarrow -n_y(x,y),\\
n_z(x,y)&\rightarrow -n_z(x,y).
\end{split}
\end{equation}
We compute the minimal set of five polarimetric measurements associated with each process (cf.~Sec.~\ref{sec:theory}), which represents the input layer of the network. To simulate experimental noise, each pixel of each polarimetric measurement is perturbed with a value extracted from a Gaussian distribution with zero mean and standard deviation ${\sigma=0.02}$. The output layer contains the parameter $\Theta(x,y)$ and the spherical representation of the vector $\bm{n}$:
\begin{equation}
\begin{split}
\theta(x,y)&=\arccos{n_z(x,y)}\\
\phi(x,y)&=\text{mod}[\text{atan2(} n_y(x,y),n_x(x,y)\text{)},2\pi],
\end{split}
\end{equation}
where $\theta(x,y)$ and $\phi(x,y)$ are the polar and azimuthal angles on the Poincaré sphere, respectively, and atan2($x,y$) is the two-argument arctangent function, which distinguishes between diametrically opposite directions.

\subsection{Neural network architecture}


To perform space-resolved process tomography, we employ a fully convolutional neural network based on the U-Net architecture. Originally used for bio-imaging segmentation~\cite{ronneberger2015u}, it has since been adapted for image-to-image regression problems in physics, such as modeling the steady-state temperature distribution from a heat source~\cite{zhao2023physics} and phase retrieval~\cite{wang2024use, proppe20243d}. Figure~\ref{fig:fig_nnconcept} outlines the structure of the architecture, which can be divided into an encoder and decoder network. The polarimetric measurements are fed into the input channels and processed through a sequence of encoding layers, wherein they are compressed into feature representations in a latent space,  
 capturing global information about the images. Crucially, such a routine is capable of recognizing the local continuity of the images. The feature representations are then fed into the decoder network, where they are eventually converted into the three ${N \times N}$ unitary parameters. 
To overcome the issue of vanishing gradients during training, skip connections send residual data from the encoder to the decoder network. We also apply Dropout regularization to prevent overfitting. Our routine can be scaled to arbitrary spatial resolutions, at the obvious cost of increasing computational complexity.

\subsection{Training strategy}
\begin{figure*}[]
    \centering
    \includegraphics[width=\linewidth]{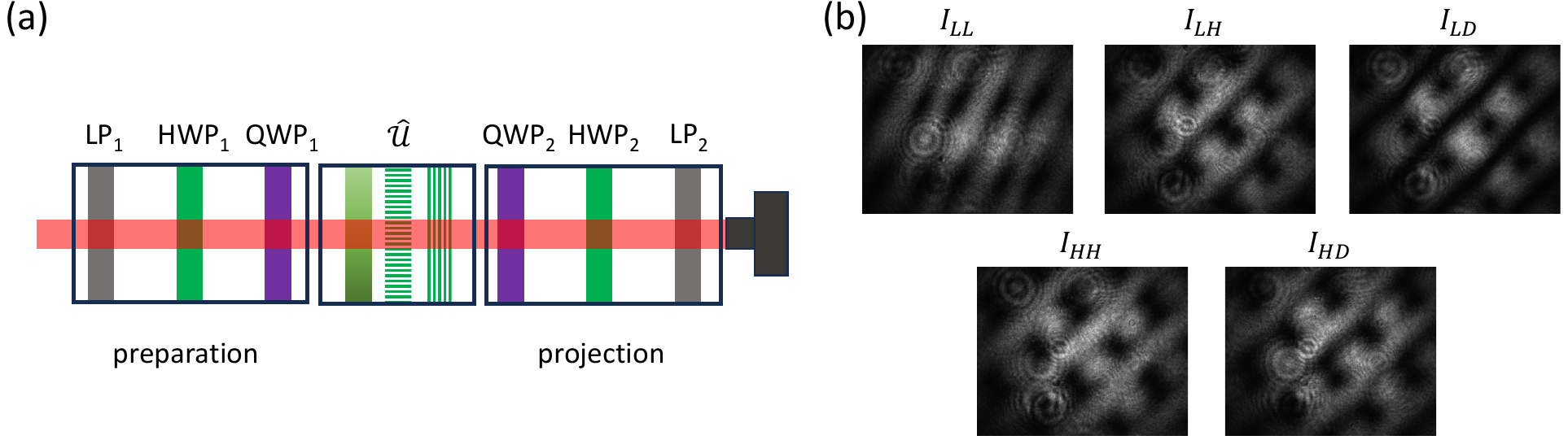}
    \caption{\textbf{Experimental process tomography.} (a) A space-dependent optical operator is implemented via LCMSs. Polarimetric measurements are realized by adjusting the preparation and projection stage. The resulting intensity distribution is collected on a camera. 
    (b) The set of five polarimetric measurements collected for the process ${U=T_{y,\Lambda}(\pi/2)T_{x,\Lambda}(\pi/2)W}$.
    }
    \label{fig:setup}
\end{figure*}

We prepare a fixed dataset of input-output training examples to perform supervised learning. The agreement between the predicted unitary process, $\hat{\mathcal{U}}_\text{exp}$, and the theoretical operator, $\hat{\mathcal{U}}_\text{th}$, is quantified in terms of the \textit{map} fidelity:

\begin{equation}
\begin{split}
    \mathcal{F} &= \frac{1}{2N^{2}}\abs{\Tr(\hat{\mathcal{U}}^{\dagger}_{\text{th}}\hat{\mathcal{U}}_{\text{exp}})}\\ &= \frac{1}{2N^{2}}\abs{\sum^{N}_{x=1}\sum^{N}_{y=1} \text{Tr}\left(U^{\dagger}_{\text{th}}(x,y)U_{\text{exp}}(x,y)\right)}.
\end{split}
    \label{eqn:mapfid}
\end{equation}
The loss function used for training is the infidelity ${1-\mathcal{F}}$. This metric constitutes the natural choice for our tomographic problem and crucially differs from the average \emph{pixel} fidelity:
\begin{equation}
 \bar{F} = \frac{1}{2N^{2}}\sum^{N}_{x=1}\sum^{N}_{y=1} \abs{\text{Tr}\left(U^{\dagger}_{\text{th}}(x,y)U_{\text{exp}}(x,y)\right)}.
\end{equation}
The latter is insensitive to relative phases between pixels. Conversely, high values of $\mathcal{F}$ certify that the routine delivers accurate predictions, also embedding the continuity constraint, which was enforced by hand in previous implementations~\cite{DiColandreaQPT}. Further details on the training hyperparameters are provided in Appendix~\ref{app:trainingDetails}.

\section{Results}

\subsection{Synthetic Experiments}
The performance of our network is first validated on $10^3$ ideal numerical experiments. These synthetic processes are generated as outlined in Sec.~\ref{main:parameterization}. 
The map fidelity (see Eq.~\eqref{eqn:mapfid}) is used to evaluate the network predictions. In view of the assessment of the network on actual experimental data, where there is no preliminary knowledge of the “true" unitary process, we also compute what we call \textit{polametric infidelity}. It quantifies the discrepancy between the input polarimetric measurements and the ones computed from the network predictions: 
\begin{equation}
    \Delta = \frac{1}{5N^2}\sum^{5}_{p=1}\sum_{x=1}^N\sum_{y=1}^N \abs{I^{(p)}_{\text{th}}\left(x,y\right) -I^{(p)}_{\text{exp}}\left(x,y\right)}^{2}, 
\end{equation}
where $p$ labels individual measurements.
Figures~\ref{fig:figure_of_merit}(a)-(b) report the distributions of the infidelity ${1-\mathcal{F}}$ and ${\Delta}$, respectively. 
The average map infidelity, ${1 - \bar{\mathcal{F}} < 0.1}$, proves that our network has been successfully trained to provide accurate reconstructions of a variety of complex processes. Correspondingly, a low value is also reported for the average polarimetric infidelity: ${\bar{\Delta}<0.05}$. 

A subset of 100 processes is then extracted to compare the performance of our network with the genetic algorithm (GA) described in Ref.~\cite{DiColandreaQPT}, by adapting the original version to the minimal case of five measurements. Figures~\ref{fig:figure_of_merit}(c)-(d) report the infidelity distribution of the two routines when processing ideal measurements and noisy data, respectively. In the second case, a Gaussian noise with  ${\sigma=0.02}$ is considered. The average performance of our routine is closely aligned with the GA. With ideal input measurements, the average infidelity is ${1-\bar{\mathcal{F}}_{CNN}=0.140}$ and ${1-\bar{\mathcal{F}}_{GA}=0.139}$ for the network and the GA, respectively. With noise, we obtain ${1-\bar{\mathcal{F}}_{CNN}=0.140}$ and ${1-\bar{\mathcal{F}}_{GA}=0.168}$. This certifies the robustness of the network to noise. Interestingly, a poor correlation between map and pixel fidelity is observed for the GA, as shown in the insets of Figs.~\ref{fig:figure_of_merit}(c)-(d). This is ascribed to the intrinsic pixel-by-pixel approach, which cannot capture the continuity of the physical parameters. The network is, in fact, immune to this effect.

On an AMD Ryzen 4500U @~2.38~GHz CPU, each reconstruction from the network takes $150$ ms on average, compared to ${\approx 60}$~s required by the GA.

\subsection{Complex polarization transformations}
Process tomography is experimentally engineered within the setup sketched in Fig.~\ref{fig:setup}(a). The source is the output of a Ti:Sa laser (central wavelength ${\lambda=810}$~nm), spatially cleaned through a single-mode fiber (not shown in the figure). Complex polarization transformations are realized via LCMSs, acting as waveplates with patterned optic-axis modulation. The birefringence of these devices is electrically controlled~\cite{Piccirillo2010}. In particular, our experiments involve different combinations of linear polarization gratings, known as $g$-plates~\cite{DErrico2020}, and plates exhibiting a uniform optic-axis orientation. In the case of ${g\text{-plates}}$ acting along $x$, $T_{x,\Lambda}(\delta)$, the optic-axis distribution is given by ${\alpha(x,y)=\pi x/\Lambda}$, with ${\Lambda=2.5}$~mm (and similarly along $y$). Uniform plates have ${\alpha(x,y)=0}$, and act as ${W=(\sigma_0+i\sigma_x)/\sqrt{2}}$. The beam waist is adjusted to ${w_0\simeq 2.5}$~mm to cover at least one spatial period of the optical operator. A linear polarizer, a half-wave plate, and a quarter-wave plate ($\text{LP}_1$-$\text{HWP}_1$-$\text{QWP}_1$) are needed to prepare an arbitrary polarization state. The polarimetric measurements are completed with a mirror sequence ($\text{QWP}_2$-$\text{HWP}_2$-$\text{LP}_2$) implementing the projection. The resulting intensity distributions are collected on a camera. The latter is placed immediately after the projection stage, so that the decomposition of Eq.~\eqref{eqn:pixeldecomposition} is verified with good approximation. This corresponds to neglecting any effect resulting from light propagation, which can be addressed via, e.g., Fourier Quantum Process Tomography~\cite{DiColandreaFQPT}.
\begin{figure*}[]
    \centering\includegraphics[width=\linewidth]{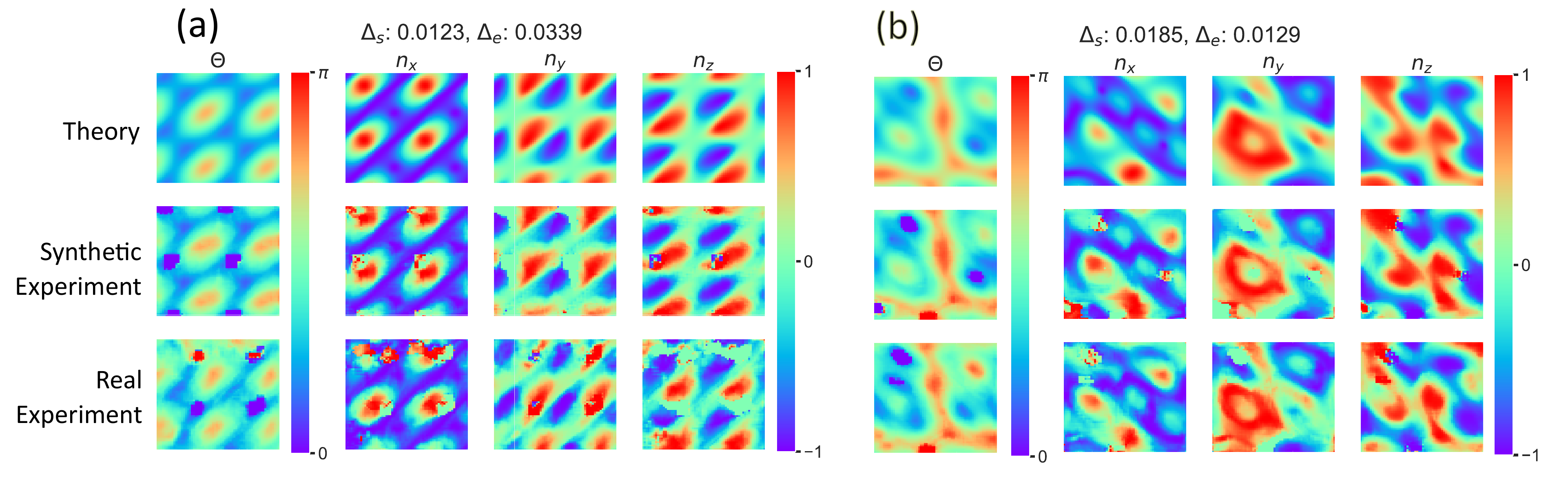}
    \caption{\textbf{Complex polarization transformations.} By combining different LCMSs, we realize the processes (a)~${T_{y,\Lambda}(\pi/2)T_{x,\Lambda}(\pi/2)W}$ and (b) ${T_{y,\Lambda/2}(1)T_{x,\Lambda/2}(1)WT_{y,\Lambda}(\pi/2)T_{x,\Lambda}(\pi/2)W}$ (see main text for the description of individual optical operators). The parameters of the theoretical process are plotted with the tomographic reconstruction resulting from both synthetic and real experimental data.}
    \label{fig:fig_exptReconstruct}
\end{figure*}
\begin{figure*}
    \centering
    \includegraphics[width=\linewidth]{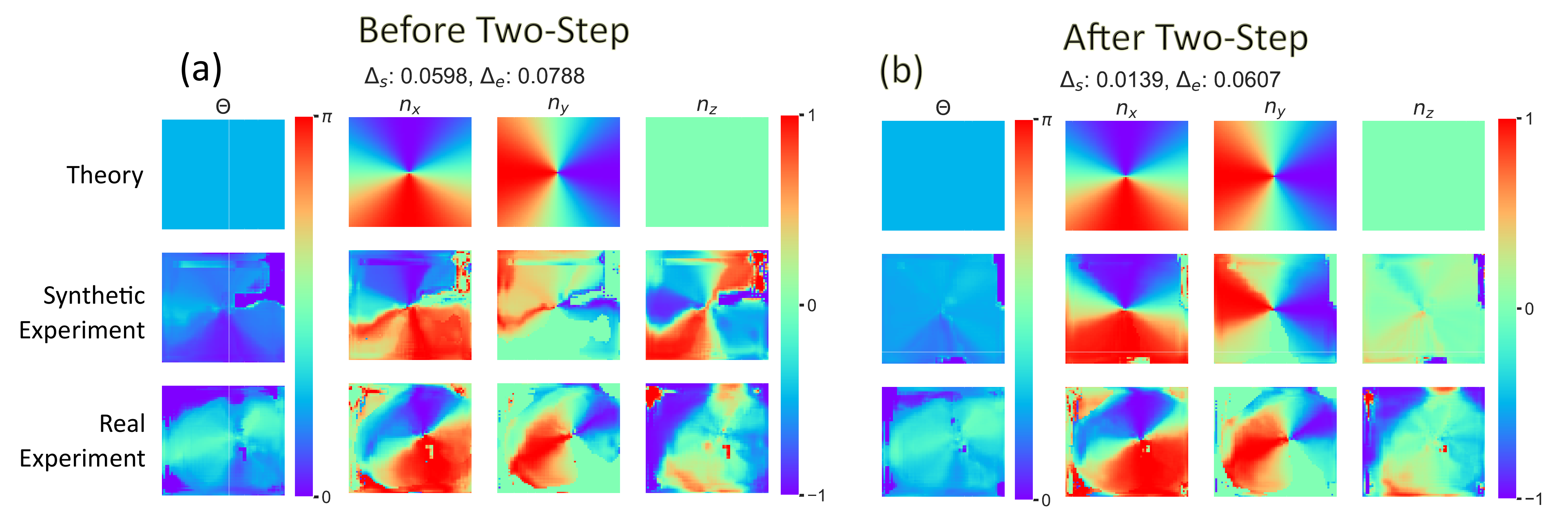}
    \caption{\textbf{Single-plate process.} We perform QPT on a $q$-plate. As expected, the first reconstruction is poor (a), but the network performance significantly improves after re-training the original architecture (b).} 
    \label{fig:fig-twoStep-qPlate}
\end{figure*}

In a first experiment, we engineer the three-plate process ${U=T_{y,\Lambda}(\pi/2)T_{x,\Lambda}(\pi/2)W}$.
Figure~\ref{fig:setup}(b) shows the intensity distributions resulting from the complete set of five polarimetric measurements. In Fig.~\ref{fig:fig_exptReconstruct}(a), we report the comparison between the theoretical unitary parameters and the network predictions on both synthetic and experimental data. The network successfully captures the distinctive features of all the modulations. The polarimetric infidelities are ${\Delta_s=0.0123}$ (corresponding to a fidelity ${\mathcal{F}=96.6\%}$) and ${\Delta_e=0.0339}$ for the synthetic and experimental reconstructions, respectively. The second experiment is realized by cascading six LCMSs as follows: ${U=T_{y,\Lambda/2}(1)T_{x,\Lambda/2}(1)WT_{y,\Lambda}(\pi/2)T_{x,\Lambda}(\pi/2)W}$. The network still provides satisfactory reconstructions, with very low polarimetric infidelities: ${\Delta_s=0.0185}$ (${\mathcal{F}=91.9\%}$) and ${\Delta_e=0.0129}$ (see Fig.~\ref{fig:fig_exptReconstruct}(b)). 
We finally prepare a third experiment reproducing a process that does not fall into the training category examples. In particular, we perform QPT on a single $q$-plate~\cite{marrucciqplate}, with ${\delta=\pi}$. Such a device is characterized by an azimuthal pattern featuring a singularity at the center: ${\alpha(x,y)=\alpha(\varphi)=q\varphi}$, where ${\varphi=\text{atan2}(y,x)}$ and $q$ is the topological charge. In our case, we set ${q=1/2}$. The network reconstruction is generally poor, as shown in Fig.~\ref{fig:fig-twoStep-qPlate}(a). This is expected for two main reasons: (i)~the unitary parameters do not feature a simple decomposition as in Eq.~\eqref{eqn:parametrizationsamples}, and (ii)~single-plate processes feature the maximal global phase ambiguity~\cite{DiColandreaQPT}, as ${n_z(x,y)=0}$. However, these problematic cases can be handled with a second stage of training, whereby the network is further trained on an augmented dataset containing a small portion of such examples. After the re-training stage, the reconstruction of the $q$-plate process significantly improves (see Fig.~\ref{fig:fig-twoStep}(b)), with $\approx80\%$ and $\approx25\%$ relative improvement on the synthetic and experimental predictions, respectively. Further details on the re-training procedure are provided in Appendix~\ref{app:specialization}, where we also investigate the general improvement of the network performance on single-plate processes.



\section{Conclusions}

The convolutional neural network we have trained to perform space-resolved process tomography is capable of outputting fast and accurate reconstructions, by parallel-processing polarimetric data and extracting global features of complex polarization transformations. We have also proven that further re-training on specialized datasets improves the performance on out-of-training examples.

Despite our implementation, these findings are not limited to optical polarization, and could be straightforwardly adapted to other fields by specializing the training set to the particular class of experiments. This study sets the baseline for the development of further optimization routines for tomographic problems in the quantum domain. Future generalizations of this work could be applied to non-unitary evolutions~\cite{Wang:23} and multi-photon gates~\cite{zhong2020quantum} in high-dimensional Hilbert spaces. 

\section*{Acknowledgements}
This work was supported by the Canada Research Chair (CRC) Program, NRC-uOttawa Joint Centre for Extreme Quantum Photonics (JCEP) via the Quantum Sensors Challenge Program at the National Research Council of Canada, and Quantum Enhanced Sensing and Imaging (QuEnSI) Alliance Consortia Quantum grant.

\clearpage
\appendix
\begin{figure*}[!t]
    \centering
    \includegraphics[width=\linewidth]{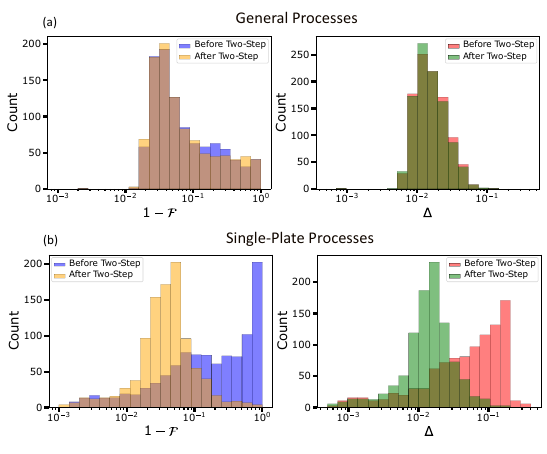}
    \caption{\textbf{Two-step training.} Distribution of the map and the polarimetric infidelity, before and after the two-step training, resulting from the network reconstruction of $10^3$ synthetic processes, generated as (a) typical training samples, and (b) single-plate processes. After the two-step training, the network performance on single-plate processes significantly improves, without attacking the performance on general processes.}
    \label{fig:fig-twoStep}
\end{figure*}

\section{Training hyperparameters}
\label{app:trainingDetails}
The complete set of hyperparameters is listed in Table~\ref{tab:tab_hyperparams}. We generate 50000 training samples with the parameterization described in Sec.~\ref{main:parameterization}. An 85:15 split between the training and the validation dataset is used. The validation stage prevents the issue of overfitting, certifying that the network is not merely adapting its parameters to the specific training set. The input consists of five ${64 \times 64}$ polarimetric images, while the unitary parameters form the output layer. Artificial noise is introduced on each pixel from a Gaussian distribution with zero mean and $\sigma=0.02$. 

\begin{table}[]
\caption{Hyperparameters.}
    \centering
    \renewcommand{\arraystretch}{1.5}
    \begin{tabular}{c | c}
    \hline
\hline
         Image Resolution& $64 \times 64$ \\ 
         Size of Dataset & $50000$ \\
         Batch Size & $8$ \\
         Train-Validation Split & 85:15 \\
         Initial Learning Rate & $10^{-3}$ \\
         Dropout Rate & $0.2$ \\
         Number of Epochs & $200$ \\
         Patience & $15$ Epochs \\ 
         Reduction Factor & $0.1$ \\
         \midrule
         Training Error at Convergence & $0.07$ \\
         Validation Error at Convergence & $0.11$ \\
         \bottomrule
         \hline
         \hline
    \end{tabular}
    \label{tab:tab_hyperparams}
\end{table}
The training process is realized within the Tensorflow library~\cite{tensorflow2015-whitepaper``}, with the help of the Adam optimizer~\cite{kingma2017adam}. We adopt an adaptive training strategy, with the learning rate that is initialized at ${10^{-3}}$ and is reduced by a factor of $10^{-1}$ if, after 15 epochs, no appreciable change in the validation loss is observed. We stopped the training after 200 epochs in correspondence with the convergence of the validation loss. Training is carried out on the \textit{Narval} supercluster using two NVidia A100SXM4 GPUs, each with 40 GB of memory. 
The process was completed in roughly 7 hours, with each epoch taking $\approx 2$ minutes.



\section{Training specialization}
\label{app:specialization}

An architecture that has been trained on a specific class of processes may struggle to reconstruct the features of more exotic examples. In our case, this occurs for single-plate optical operators.
This limitation can be overcome by re-training the original network on a specialized dataset containing a portion of problematic examples, in reminiscence of fine-tuning Language Learning Models towards desired properties~\cite{taylor2022galactica}. This is known as Transfer Learning~\cite{TransferLearning}. Specifically, this is accomplished by augmenting the original dataset with $6000$ examples consisting of single- and two-plate processes. For each waveplate, the birefringence $\delta$ is uniformly extracted from the range $[0, 2\pi[$, while the optic-axis orientation follows a decomposition similar to Eq.~\eqref{eqn:parametrizationsamples}:
\begin{equation}
\begin{split}
\alpha(x,y)=\sum_{i=0}^{\Omega_x}\sum_{j=0}^{\Omega_y}c^{(1)}_{ij}\cos{\dfrac{2\pi i x}{N}}\cos{\dfrac{2\pi j y}{N}}+&\\c^{(2)}_{ij}\cos{\dfrac{2\pi i x}{N}}\sin{\dfrac{2\pi j y}{N}}+&\\c^{(3)}_{ij}\sin{\dfrac{2\pi i x}{N}}\cos{\dfrac{2\pi j y}{N}}+&\\c^{(4)}_{ij}\sin{\dfrac{2\pi i x}{N}}\sin{\dfrac{2\pi j y}{N}},
\label{eqn:suppeq}
\end{split}
\end{equation}
with ${0\leq \Omega_x\leq 3}$, ${0\leq \Omega_y\leq 3}$, and each coefficient ${c^{(m)}_{ij}}$ is uniformly extracted from the range ${[-1,1]}$. 

Importantly, the ratio of these examples with the original dataset is kept reasonably small to avoid biasing the network towards these processes. Indeed, the ability of our network to reconstruct general processes is not diminished after Transfer Learning, as shown in Fig.~\ref{fig:fig-twoStep}(a). Conversely, the network demonstrates improved performance in predicting complex single-plate processes, as can be seen in Fig.~\ref{fig:fig-twoStep}(b) by the shift of the initial infidelity distributions towards smaller values.


\newpage
\clearpage
\bibliography{main}

\begin{thebibliography}{45}%
\makeatletter
\providecommand \@ifxundefined [1]{%
 \@ifx{#1\undefined}
}%
\providecommand \@ifnum [1]{%
 \ifnum #1\expandafter \@firstoftwo
 \else \expandafter \@secondoftwo
 \fi
}%
\providecommand \@ifx [1]{%
 \ifx #1\expandafter \@firstoftwo
 \else \expandafter \@secondoftwo
 \fi
}%
\providecommand \natexlab [1]{#1}%
\providecommand \enquote  [1]{``#1''}%
\providecommand \bibnamefont  [1]{#1}%
\providecommand \bibfnamefont [1]{#1}%
\providecommand \citenamefont [1]{#1}%
\providecommand \href@noop [0]{\@secondoftwo}%
\providecommand \href [0]{\begingroup \@sanitize@url \@href}%
\providecommand \@href[1]{\@@startlink{#1}\@@href}%
\providecommand \@@href[1]{\endgroup#1\@@endlink}%
\providecommand \@sanitize@url [0]{\catcode `\\12\catcode `\$12\catcode
  `\&12\catcode `\#12\catcode `\^12\catcode `\_12\catcode `\%12\relax}%
\providecommand \@@startlink[1]{}%
\providecommand \@@endlink[0]{}%
\providecommand \url  [0]{\begingroup\@sanitize@url \@url }%
\providecommand \@url [1]{\endgroup\@href {#1}{\urlprefix }}%
\providecommand \urlprefix  [0]{URL }%
\providecommand \Eprint [0]{\href }%
\providecommand \doibase [0]{https://doi.org/}%
\providecommand \selectlanguage [0]{\@gobble}%
\providecommand \bibinfo  [0]{\@secondoftwo}%
\providecommand \bibfield  [0]{\@secondoftwo}%
\providecommand \translation [1]{[#1]}%
\providecommand \BibitemOpen [0]{}%
\providecommand \bibitemStop [0]{}%
\providecommand \bibitemNoStop [0]{.\EOS\space}%
\providecommand \EOS [0]{\spacefactor3000\relax}%
\providecommand \BibitemShut  [1]{\csname bibitem#1\endcsname}%
\let\auto@bib@innerbib\@empty
\bibitem [{\citenamefont {Chuang}\ and\ \citenamefont
  {Nielsen}(1997)}]{Chuang1997}%
  \BibitemOpen
  \bibfield  {author} {\bibinfo {author} {\bibfnamefont {I.~L.}\ \bibnamefont
  {Chuang}}\ and\ \bibinfo {author} {\bibfnamefont {M.~A.}\ \bibnamefont
  {Nielsen}},\ }\bibfield  {title} {\bibinfo {title} {{Prescription for
  experimental determination of the dynamics of a quantum black box}},\ }\href
  {https://doi.org/10.1080/09500349708231894} {\bibfield  {journal} {\bibinfo
  {journal} {J. Mod. Opt.}\ }\textbf {\bibinfo {volume} {44}},\ \bibinfo
  {pages} {2455} (\bibinfo {year} {1997})}\BibitemShut {NoStop}%
\bibitem [{\citenamefont {S{\"o}derstr{\"o}m}\ and\ \citenamefont
  {Stoica}(1989)}]{stoica}%
  \BibitemOpen
  \bibfield  {author} {\bibinfo {author} {\bibfnamefont {T.}~\bibnamefont
  {S{\"o}derstr{\"o}m}}\ and\ \bibinfo {author} {\bibfnamefont
  {P.}~\bibnamefont {Stoica}},\ }\href
  {https://books.google.ca/books?id=BClVPgAACAAJ} {\emph {\bibinfo {title}
  {System Identification}}},\ Prentice-Hall International Series in Systems and
  Control Engineering\ (\bibinfo  {publisher} {Prentice Hall},\ \bibinfo {year}
  {1989})\BibitemShut {NoStop}%
\bibitem [{\citenamefont {Childs}\ \emph {et~al.}(2001)\citenamefont {Childs},
  \citenamefont {Chuang},\ and\ \citenamefont {Leung}}]{PhysRevA.64.012314}%
  \BibitemOpen
  \bibfield  {author} {\bibinfo {author} {\bibfnamefont {A.~M.}\ \bibnamefont
  {Childs}}, \bibinfo {author} {\bibfnamefont {I.~L.}\ \bibnamefont {Chuang}},\
  and\ \bibinfo {author} {\bibfnamefont {D.~W.}\ \bibnamefont {Leung}},\
  }\bibfield  {title} {\bibinfo {title} {{Realization of quantum process
  tomography in NMR}},\ }\href {https://doi.org/10.1103/PhysRevA.64.012314}
  {\bibfield  {journal} {\bibinfo  {journal} {Phys. Rev. A}\ }\textbf {\bibinfo
  {volume} {64}},\ \bibinfo {pages} {012314} (\bibinfo {year}
  {2001})}\BibitemShut {NoStop}%
\bibitem [{\citenamefont {Myrskog}\ \emph {et~al.}(2005)\citenamefont
  {Myrskog}, \citenamefont {Fox}, \citenamefont {Mitchell},\ and\ \citenamefont
  {Steinberg}}]{PhysRevA.72.013615}%
  \BibitemOpen
  \bibfield  {author} {\bibinfo {author} {\bibfnamefont {S.~H.}\ \bibnamefont
  {Myrskog}}, \bibinfo {author} {\bibfnamefont {J.~K.}\ \bibnamefont {Fox}},
  \bibinfo {author} {\bibfnamefont {M.~W.}\ \bibnamefont {Mitchell}},\ and\
  \bibinfo {author} {\bibfnamefont {A.~M.}\ \bibnamefont {Steinberg}},\
  }\bibfield  {title} {\bibinfo {title} {Quantum process tomography on
  vibrational states of atoms in an optical lattice},\ }\href
  {https://doi.org/10.1103/PhysRevA.72.013615} {\bibfield  {journal} {\bibinfo
  {journal} {Phys. Rev. A}\ }\textbf {\bibinfo {volume} {72}},\ \bibinfo
  {pages} {013615} (\bibinfo {year} {2005})}\BibitemShut {NoStop}%
\bibitem [{\citenamefont {Roos}\ \emph {et~al.}(2004)\citenamefont {Roos},
  \citenamefont {Lancaster}, \citenamefont {Riebe}, \citenamefont {H\"affner},
  \citenamefont {H\"ansel}, \citenamefont {Gulde}, \citenamefont {Becher},
  \citenamefont {Eschner}, \citenamefont {Schmidt-Kaler},\ and\ \citenamefont
  {Blatt}}]{PhysRevLett.92.220402}%
  \BibitemOpen
  \bibfield  {author} {\bibinfo {author} {\bibfnamefont {C.~F.}\ \bibnamefont
  {Roos}}, \bibinfo {author} {\bibfnamefont {G.~P.~T.}\ \bibnamefont
  {Lancaster}}, \bibinfo {author} {\bibfnamefont {M.}~\bibnamefont {Riebe}},
  \bibinfo {author} {\bibfnamefont {H.}~\bibnamefont {H\"affner}}, \bibinfo
  {author} {\bibfnamefont {W.}~\bibnamefont {H\"ansel}}, \bibinfo {author}
  {\bibfnamefont {S.}~\bibnamefont {Gulde}}, \bibinfo {author} {\bibfnamefont
  {C.}~\bibnamefont {Becher}}, \bibinfo {author} {\bibfnamefont
  {J.}~\bibnamefont {Eschner}}, \bibinfo {author} {\bibfnamefont
  {F.}~\bibnamefont {Schmidt-Kaler}},\ and\ \bibinfo {author} {\bibfnamefont
  {R.}~\bibnamefont {Blatt}},\ }\bibfield  {title} {\bibinfo {title} {Bell
  states of atoms with ultralong lifetimes and their tomographic state
  analysis},\ }\href
  {https://doi.org/http://dx.doi.org/10.1103/PhysRevLett.92.220402} {\bibfield
  {journal} {\bibinfo  {journal} {Phys. Rev. Lett.}\ }\textbf {\bibinfo
  {volume} {92}},\ \bibinfo {pages} {220402} (\bibinfo {year}
  {2004})}\BibitemShut {NoStop}%
\bibitem [{\citenamefont {Riebe}\ \emph {et~al.}(2006)\citenamefont {Riebe},
  \citenamefont {Kim}, \citenamefont {Schindler}, \citenamefont {Monz},
  \citenamefont {Schmidt}, \citenamefont {K\"orber}, \citenamefont {H\"ansel},
  \citenamefont {H\"affner}, \citenamefont {Roos},\ and\ \citenamefont
  {Blatt}}]{PhysRevLett.97.220407}%
  \BibitemOpen
  \bibfield  {author} {\bibinfo {author} {\bibfnamefont {M.}~\bibnamefont
  {Riebe}}, \bibinfo {author} {\bibfnamefont {K.}~\bibnamefont {Kim}}, \bibinfo
  {author} {\bibfnamefont {P.}~\bibnamefont {Schindler}}, \bibinfo {author}
  {\bibfnamefont {T.}~\bibnamefont {Monz}}, \bibinfo {author} {\bibfnamefont
  {P.~O.}\ \bibnamefont {Schmidt}}, \bibinfo {author} {\bibfnamefont {T.~K.}\
  \bibnamefont {K\"orber}}, \bibinfo {author} {\bibfnamefont {W.}~\bibnamefont
  {H\"ansel}}, \bibinfo {author} {\bibfnamefont {H.}~\bibnamefont {H\"affner}},
  \bibinfo {author} {\bibfnamefont {C.~F.}\ \bibnamefont {Roos}},\ and\
  \bibinfo {author} {\bibfnamefont {R.}~\bibnamefont {Blatt}},\ }\bibfield
  {title} {\bibinfo {title} {Process tomography of ion trap quantum gates},\
  }\href {https://doi.org/10.1103/PhysRevLett.97.220407} {\bibfield  {journal}
  {\bibinfo  {journal} {Phys. Rev. Lett.}\ }\textbf {\bibinfo {volume} {97}},\
  \bibinfo {pages} {220407} (\bibinfo {year} {2006})}\BibitemShut {NoStop}%
\bibitem [{\citenamefont {Mitchell}\ \emph {et~al.}(2003)\citenamefont
  {Mitchell}, \citenamefont {Ellenor}, \citenamefont {Schneider},\ and\
  \citenamefont {Steinberg}}]{PhysRevLett.91.120402}%
  \BibitemOpen
  \bibfield  {author} {\bibinfo {author} {\bibfnamefont {M.~W.}\ \bibnamefont
  {Mitchell}}, \bibinfo {author} {\bibfnamefont {C.~W.}\ \bibnamefont
  {Ellenor}}, \bibinfo {author} {\bibfnamefont {S.}~\bibnamefont {Schneider}},\
  and\ \bibinfo {author} {\bibfnamefont {A.~M.}\ \bibnamefont {Steinberg}},\
  }\bibfield  {title} {\bibinfo {title} {Diagnosis, prescription, and prognosis
  of a bell-state filter by quantum process tomography},\ }\href
  {https://doi.org/10.1103/PhysRevLett.91.120402} {\bibfield  {journal}
  {\bibinfo  {journal} {Phys. Rev. Lett.}\ }\textbf {\bibinfo {volume} {91}},\
  \bibinfo {pages} {120402} (\bibinfo {year} {2003})}\BibitemShut {NoStop}%
\bibitem [{\citenamefont {Altepeter}\ \emph {et~al.}(2003)\citenamefont
  {Altepeter}, \citenamefont {Branning}, \citenamefont {Jeffrey}, \citenamefont
  {Wei}, \citenamefont {Kwiat}, \citenamefont {Thew}, \citenamefont {O'Brien},
  \citenamefont {Nielsen},\ and\ \citenamefont {White}}]{Altepeter2003}%
  \BibitemOpen
  \bibfield  {author} {\bibinfo {author} {\bibfnamefont {J.~B.}\ \bibnamefont
  {Altepeter}}, \bibinfo {author} {\bibfnamefont {D.}~\bibnamefont {Branning}},
  \bibinfo {author} {\bibfnamefont {E.}~\bibnamefont {Jeffrey}}, \bibinfo
  {author} {\bibfnamefont {T.~C.}\ \bibnamefont {Wei}}, \bibinfo {author}
  {\bibfnamefont {P.~G.}\ \bibnamefont {Kwiat}}, \bibinfo {author}
  {\bibfnamefont {R.~T.}\ \bibnamefont {Thew}}, \bibinfo {author}
  {\bibfnamefont {J.~L.}\ \bibnamefont {O'Brien}}, \bibinfo {author}
  {\bibfnamefont {M.~A.}\ \bibnamefont {Nielsen}},\ and\ \bibinfo {author}
  {\bibfnamefont {A.~G.}\ \bibnamefont {White}},\ }\bibfield  {title} {\bibinfo
  {title} {{Ancilla-Assisted Quantum Process Tomography}},\ }\href
  {https://doi.org/https://dx.doi.org/10.1103/PhysRevLett.90.193601} {\bibfield
   {journal} {\bibinfo  {journal} {Phys. Rev. Lett.}\ }\textbf {\bibinfo
  {volume} {90}},\ \bibinfo {pages} {193601} (\bibinfo {year}
  {2003})}\BibitemShut {NoStop}%
\bibitem [{\citenamefont {O'Brien}\ \emph {et~al.}(2004)\citenamefont
  {O'Brien}, \citenamefont {Pryde}, \citenamefont {Gilchrist}, \citenamefont
  {James}, \citenamefont {Langford}, \citenamefont {Ralph},\ and\ \citenamefont
  {White}}]{PhysRevLett.93.080502}%
  \BibitemOpen
  \bibfield  {author} {\bibinfo {author} {\bibfnamefont {J.~L.}\ \bibnamefont
  {O'Brien}}, \bibinfo {author} {\bibfnamefont {G.~J.}\ \bibnamefont {Pryde}},
  \bibinfo {author} {\bibfnamefont {A.}~\bibnamefont {Gilchrist}}, \bibinfo
  {author} {\bibfnamefont {D.~F.~V.}\ \bibnamefont {James}}, \bibinfo {author}
  {\bibfnamefont {N.~K.}\ \bibnamefont {Langford}}, \bibinfo {author}
  {\bibfnamefont {T.~C.}\ \bibnamefont {Ralph}},\ and\ \bibinfo {author}
  {\bibfnamefont {A.~G.}\ \bibnamefont {White}},\ }\bibfield  {title} {\bibinfo
  {title} {Quantum process tomography of a controlled-not gate},\ }\href
  {https://doi.org/10.1103/PhysRevLett.93.080502} {\bibfield  {journal}
  {\bibinfo  {journal} {Phys. Rev. Lett.}\ }\textbf {\bibinfo {volume} {93}},\
  \bibinfo {pages} {080502} (\bibinfo {year} {2004})}\BibitemShut {NoStop}%
\bibitem [{\citenamefont {Lobino}\ \emph {et~al.}(2008)\citenamefont {Lobino},
  \citenamefont {Korystov}, \citenamefont {Kupchak}, \citenamefont {Figueroa},
  \citenamefont {Sanders},\ and\ \citenamefont {Lvovsky}}]{Lobino2008}%
  \BibitemOpen
  \bibfield  {author} {\bibinfo {author} {\bibfnamefont {M.}~\bibnamefont
  {Lobino}}, \bibinfo {author} {\bibfnamefont {D.}~\bibnamefont {Korystov}},
  \bibinfo {author} {\bibfnamefont {C.}~\bibnamefont {Kupchak}}, \bibinfo
  {author} {\bibfnamefont {E.}~\bibnamefont {Figueroa}}, \bibinfo {author}
  {\bibfnamefont {B.~C.}\ \bibnamefont {Sanders}},\ and\ \bibinfo {author}
  {\bibfnamefont {A.~I.}\ \bibnamefont {Lvovsky}},\ }\bibfield  {title}
  {\bibinfo {title} {{Complete Characterization of Quantum-Optical
  Processes}},\ }\href
  {https://doi.org/http://dx.doi.org/10.1126/science.1162086} {\bibfield
  {journal} {\bibinfo  {journal} {Science}\ }\textbf {\bibinfo {volume}
  {322}},\ \bibinfo {pages} {563} (\bibinfo {year} {2008})}\BibitemShut
  {NoStop}%
\bibitem [{\citenamefont {Bongioanni}\ \emph {et~al.}(2010)\citenamefont
  {Bongioanni}, \citenamefont {Sansoni}, \citenamefont {Sciarrino},
  \citenamefont {Vallone},\ and\ \citenamefont {Mataloni}}]{Bongioanni2010}%
  \BibitemOpen
  \bibfield  {author} {\bibinfo {author} {\bibfnamefont {I.}~\bibnamefont
  {Bongioanni}}, \bibinfo {author} {\bibfnamefont {L.}~\bibnamefont {Sansoni}},
  \bibinfo {author} {\bibfnamefont {F.}~\bibnamefont {Sciarrino}}, \bibinfo
  {author} {\bibfnamefont {G.}~\bibnamefont {Vallone}},\ and\ \bibinfo {author}
  {\bibfnamefont {P.}~\bibnamefont {Mataloni}},\ }\bibfield  {title} {\bibinfo
  {title} {{Experimental quantum process tomography of non-trace-preserving
  maps}},\ }\href {https://doi.org/10.1103/PhysRevA.82.042307} {\bibfield
  {journal} {\bibinfo  {journal} {Phys. Rev. A}\ }\textbf {\bibinfo {volume}
  {82}},\ \bibinfo {pages} {042307} (\bibinfo {year} {2010})}\BibitemShut
  {NoStop}%
\bibitem [{\citenamefont {Rahimi-Keshari}\ \emph {et~al.}(2013)\citenamefont
  {Rahimi-Keshari}, \citenamefont {Broome}, \citenamefont {Fickler},
  \citenamefont {Fedrizzi}, \citenamefont {Ralph},\ and\ \citenamefont
  {White}}]{Rahimi-Keshari2013}%
  \BibitemOpen
  \bibfield  {author} {\bibinfo {author} {\bibfnamefont {S.}~\bibnamefont
  {Rahimi-Keshari}}, \bibinfo {author} {\bibfnamefont {M.~A.}\ \bibnamefont
  {Broome}}, \bibinfo {author} {\bibfnamefont {R.}~\bibnamefont {Fickler}},
  \bibinfo {author} {\bibfnamefont {A.}~\bibnamefont {Fedrizzi}}, \bibinfo
  {author} {\bibfnamefont {T.~C.}\ \bibnamefont {Ralph}},\ and\ \bibinfo
  {author} {\bibfnamefont {A.~G.}\ \bibnamefont {White}},\ }\bibfield  {title}
  {\bibinfo {title} {{Direct characterization of linear-optical networks}},\
  }\href {https://doi.org/http://dx.doi.org/10.1364/OE.21.013450} {\bibfield
  {journal} {\bibinfo  {journal} {Opt. Express}\ }\textbf {\bibinfo {volume}
  {21}},\ \bibinfo {pages} {13450} (\bibinfo {year} {2013})}\BibitemShut
  {NoStop}%
\bibitem [{\citenamefont {Zhou}\ \emph {et~al.}(2015)\citenamefont {Zhou},
  \citenamefont {Cable}, \citenamefont {Whittaker}, \citenamefont {Shadbolt},
  \citenamefont {O'Brien},\ and\ \citenamefont {Matthews}}]{Zhou:15}%
  \BibitemOpen
  \bibfield  {author} {\bibinfo {author} {\bibfnamefont {X.-Q.}\ \bibnamefont
  {Zhou}}, \bibinfo {author} {\bibfnamefont {H.}~\bibnamefont {Cable}},
  \bibinfo {author} {\bibfnamefont {R.}~\bibnamefont {Whittaker}}, \bibinfo
  {author} {\bibfnamefont {P.}~\bibnamefont {Shadbolt}}, \bibinfo {author}
  {\bibfnamefont {J.~L.}\ \bibnamefont {O'Brien}},\ and\ \bibinfo {author}
  {\bibfnamefont {J.~C.~F.}\ \bibnamefont {Matthews}},\ }\bibfield  {title}
  {\bibinfo {title} {Quantum-enhanced tomography of unitary processes},\ }\href
  {https://doi.org/10.1364/OPTICA.2.000510} {\bibfield  {journal} {\bibinfo
  {journal} {Optica}\ }\textbf {\bibinfo {volume} {2}},\ \bibinfo {pages} {510}
  (\bibinfo {year} {2015})}\BibitemShut {NoStop}%
\bibitem [{\citenamefont {Ant{\'{o}}n}\ \emph {et~al.}(2017)\citenamefont
  {Ant{\'{o}}n}, \citenamefont {Hilaire}, \citenamefont {Kessler},
  \citenamefont {Demory}, \citenamefont {G{\'{o}}mez}, \citenamefont
  {Lema{\^{i}}tre}, \citenamefont {Sagnes}, \citenamefont {Lanzillotti-Kimura},
  \citenamefont {Krebs}, \citenamefont {Somaschi}, \citenamefont {Senellart},\
  and\ \citenamefont {Lanco}}]{Anton2017}%
  \BibitemOpen
  \bibfield  {author} {\bibinfo {author} {\bibfnamefont {C.}~\bibnamefont
  {Ant{\'{o}}n}}, \bibinfo {author} {\bibfnamefont {P.}~\bibnamefont
  {Hilaire}}, \bibinfo {author} {\bibfnamefont {C.~A.}\ \bibnamefont
  {Kessler}}, \bibinfo {author} {\bibfnamefont {J.}~\bibnamefont {Demory}},
  \bibinfo {author} {\bibfnamefont {C.}~\bibnamefont {G{\'{o}}mez}}, \bibinfo
  {author} {\bibfnamefont {A.}~\bibnamefont {Lema{\^{i}}tre}}, \bibinfo
  {author} {\bibfnamefont {I.}~\bibnamefont {Sagnes}}, \bibinfo {author}
  {\bibfnamefont {N.~D.}\ \bibnamefont {Lanzillotti-Kimura}}, \bibinfo {author}
  {\bibfnamefont {O.}~\bibnamefont {Krebs}}, \bibinfo {author} {\bibfnamefont
  {N.}~\bibnamefont {Somaschi}}, \bibinfo {author} {\bibfnamefont
  {P.}~\bibnamefont {Senellart}},\ and\ \bibinfo {author} {\bibfnamefont
  {L.}~\bibnamefont {Lanco}},\ }\bibfield  {title} {\bibinfo {title}
  {{Tomography of the optical polarization rotation induced by a single quantum
  dot in a cavity}},\ }\href {https://doi.org/10.1364/OPTICA.4.001326}
  {\bibfield  {journal} {\bibinfo  {journal} {Optica}\ }\textbf {\bibinfo
  {volume} {4}},\ \bibinfo {pages} {1326} (\bibinfo {year} {2017})}\BibitemShut
  {NoStop}%
\bibitem [{\citenamefont {Jacob}\ \emph {et~al.}(2018)\citenamefont {Jacob},
  \citenamefont {Mirasola}, \citenamefont {Adhikari},\ and\ \citenamefont
  {Dowling}}]{PhysRevA.98.052327}%
  \BibitemOpen
  \bibfield  {author} {\bibinfo {author} {\bibfnamefont {K.~V.}\ \bibnamefont
  {Jacob}}, \bibinfo {author} {\bibfnamefont {A.~E.}\ \bibnamefont {Mirasola}},
  \bibinfo {author} {\bibfnamefont {S.}~\bibnamefont {Adhikari}},\ and\
  \bibinfo {author} {\bibfnamefont {J.~P.}\ \bibnamefont {Dowling}},\
  }\bibfield  {title} {\bibinfo {title} {Direct characterization of linear and
  quadratically nonlinear optical systems},\ }\href
  {https://doi.org/10.1103/PhysRevA.98.052327} {\bibfield  {journal} {\bibinfo
  {journal} {Phys. Rev. A}\ }\textbf {\bibinfo {volume} {98}},\ \bibinfo
  {pages} {052327} (\bibinfo {year} {2018})}\BibitemShut {NoStop}%
\bibitem [{\citenamefont {Bouchard}\ \emph {et~al.}(2019)\citenamefont
  {Bouchard}, \citenamefont {Hufnagel}, \citenamefont {Koutn{\'{y}}},
  \citenamefont {Abbas}, \citenamefont {Sit}, \citenamefont {Heshami},
  \citenamefont {Fickler},\ and\ \citenamefont {Karimi}}]{Bouchard2019}%
  \BibitemOpen
  \bibfield  {author} {\bibinfo {author} {\bibfnamefont {F.}~\bibnamefont
  {Bouchard}}, \bibinfo {author} {\bibfnamefont {F.}~\bibnamefont {Hufnagel}},
  \bibinfo {author} {\bibfnamefont {D.}~\bibnamefont {Koutn{\'{y}}}}, \bibinfo
  {author} {\bibfnamefont {A.}~\bibnamefont {Abbas}}, \bibinfo {author}
  {\bibfnamefont {A.}~\bibnamefont {Sit}}, \bibinfo {author} {\bibfnamefont
  {K.}~\bibnamefont {Heshami}}, \bibinfo {author} {\bibfnamefont
  {R.}~\bibnamefont {Fickler}},\ and\ \bibinfo {author} {\bibfnamefont
  {E.}~\bibnamefont {Karimi}},\ }\bibfield  {title} {\bibinfo {title} {{Quantum
  process tomography of a high-dimensional quantum communication channel}},\
  }\href {https://doi.org/http://dx.doi.org/10.22331/q-2019-05-06-138}
  {\bibfield  {journal} {\bibinfo  {journal} {Quantum}\ }\textbf {\bibinfo
  {volume} {3}},\ \bibinfo {pages} {138} (\bibinfo {year} {2019})}\BibitemShut
  {NoStop}%
\bibitem [{\citenamefont {{Di Colandrea}}\ \emph
  {et~al.}(2023{\natexlab{a}})\citenamefont {{Di Colandrea}}, \citenamefont
  {Amato}, \citenamefont {Schiattarella}, \citenamefont {Dauphin},\ and\
  \citenamefont {Cardano}}]{DiColandreaQPT}%
  \BibitemOpen
  \bibfield  {author} {\bibinfo {author} {\bibfnamefont {F.}~\bibnamefont {{Di
  Colandrea}}}, \bibinfo {author} {\bibfnamefont {L.}~\bibnamefont {Amato}},
  \bibinfo {author} {\bibfnamefont {R.}~\bibnamefont {Schiattarella}}, \bibinfo
  {author} {\bibfnamefont {A.}~\bibnamefont {Dauphin}},\ and\ \bibinfo {author}
  {\bibfnamefont {F.}~\bibnamefont {Cardano}},\ }\bibfield  {title} {\bibinfo
  {title} {Retrieving space-dependent polarization transformations via
  near-optimal quantum process tomography},\ }\href
  {https://doi.org/10.1364/OE.491518} {\bibfield  {journal} {\bibinfo
  {journal} {Opt. Express}\ }\textbf {\bibinfo {volume} {31}},\ \bibinfo
  {pages} {31698} (\bibinfo {year} {2023}{\natexlab{a}})}\BibitemShut {NoStop}%
\bibitem [{\citenamefont {Goel}\ \emph {et~al.}(2024)\citenamefont {Goel},
  \citenamefont {Leedumrongwatthanakun}, \citenamefont {Valencia},
  \citenamefont {McCutcheon}, \citenamefont {Tavakoli}, \citenamefont {Conti},
  \citenamefont {Pinkse},\ and\ \citenamefont {Malik}}]{goel2024inverse}%
  \BibitemOpen
  \bibfield  {author} {\bibinfo {author} {\bibfnamefont {S.}~\bibnamefont
  {Goel}}, \bibinfo {author} {\bibfnamefont {S.}~\bibnamefont
  {Leedumrongwatthanakun}}, \bibinfo {author} {\bibfnamefont {N.~H.}\
  \bibnamefont {Valencia}}, \bibinfo {author} {\bibfnamefont {W.}~\bibnamefont
  {McCutcheon}}, \bibinfo {author} {\bibfnamefont {A.}~\bibnamefont
  {Tavakoli}}, \bibinfo {author} {\bibfnamefont {C.}~\bibnamefont {Conti}},
  \bibinfo {author} {\bibfnamefont {P.~W.}\ \bibnamefont {Pinkse}},\ and\
  \bibinfo {author} {\bibfnamefont {M.}~\bibnamefont {Malik}},\ }\bibfield
  {title} {\bibinfo {title} {Inverse design of high-dimensional quantum optical
  circuits in a complex medium},\ }\href
  {https://www.nature.com/articles/s41567-023-02319-6} {\bibfield  {journal}
  {\bibinfo  {journal} {Nat. Phys.}\ }\textbf {\bibinfo {volume} {20}},\
  \bibinfo {pages} {1} (\bibinfo {year} {2024})}\BibitemShut {NoStop}%
\bibitem [{\citenamefont {James}\ \emph {et~al.}(2001)\citenamefont {James},
  \citenamefont {Kwiat}, \citenamefont {Munro},\ and\ \citenamefont
  {White}}]{James2001}%
  \BibitemOpen
  \bibfield  {author} {\bibinfo {author} {\bibfnamefont {D.~F.~V.}\
  \bibnamefont {James}}, \bibinfo {author} {\bibfnamefont {P.~G.}\ \bibnamefont
  {Kwiat}}, \bibinfo {author} {\bibfnamefont {W.~J.}\ \bibnamefont {Munro}},\
  and\ \bibinfo {author} {\bibfnamefont {A.~G.}\ \bibnamefont {White}},\
  }\bibfield  {title} {\bibinfo {title} {Measurement of qubits},\ }\href
  {https://doi.org/10.1103/PhysRevA.64.052312} {\bibfield  {journal} {\bibinfo
  {journal} {Phys. Rev. A}\ }\textbf {\bibinfo {volume} {64}},\ \bibinfo
  {pages} {052312} (\bibinfo {year} {2001})}\BibitemShut {NoStop}%
\bibitem [{\citenamefont {Aiello}\ \emph {et~al.}(2006)\citenamefont {Aiello},
  \citenamefont {Puentes}, \citenamefont {Voigt},\ and\ \citenamefont
  {Woerdman}}]{Aiello2006}%
  \BibitemOpen
  \bibfield  {author} {\bibinfo {author} {\bibfnamefont {A.}~\bibnamefont
  {Aiello}}, \bibinfo {author} {\bibfnamefont {G.}~\bibnamefont {Puentes}},
  \bibinfo {author} {\bibfnamefont {D.}~\bibnamefont {Voigt}},\ and\ \bibinfo
  {author} {\bibfnamefont {J.~P.}\ \bibnamefont {Woerdman}},\ }\bibfield
  {title} {\bibinfo {title} {{Maximum-likelihood estimation of Mueller
  matrices}},\ }\href {https://doi.org/https://dx.doi.org/10.1364/OL.31.000817}
  {\bibfield  {journal} {\bibinfo  {journal} {Opt. Lett.}\ }\textbf {\bibinfo
  {volume} {31}},\ \bibinfo {pages} {817} (\bibinfo {year} {2006})}\BibitemShut
  {NoStop}%
\bibitem [{\citenamefont {Solomon}(1981)}]{Solomon1981}%
  \BibitemOpen
  \bibfield  {author} {\bibinfo {author} {\bibfnamefont {J.~E.}\ \bibnamefont
  {Solomon}},\ }\bibfield  {title} {\bibinfo {title} {{Polarization imaging}},\
  }\href {https://doi.org/https://dx.doi.org/10.1364/AO.20.001537} {\bibfield
  {journal} {\bibinfo  {journal} {Appl. Opt.}\ }\textbf {\bibinfo {volume}
  {20}},\ \bibinfo {pages} {1537} (\bibinfo {year} {1981})}\BibitemShut
  {NoStop}%
\bibitem [{\citenamefont {Davis}\ \emph {et~al.}(2005)\citenamefont {Davis},
  \citenamefont {Evans},\ and\ \citenamefont {Moreno}}]{Davis2005}%
  \BibitemOpen
  \bibfield  {author} {\bibinfo {author} {\bibfnamefont {J.~A.}\ \bibnamefont
  {Davis}}, \bibinfo {author} {\bibfnamefont {G.~H.}\ \bibnamefont {Evans}},\
  and\ \bibinfo {author} {\bibfnamefont {I.}~\bibnamefont {Moreno}},\
  }\bibfield  {title} {\bibinfo {title} {{Polarization-multiplexed diffractive
  optical elements with liquid-crystal displays}},\ }\href
  {https://doi.org/https://dx.doi.org/10.1364/AO.44.004049} {\bibfield
  {journal} {\bibinfo  {journal} {Appl. Opt.}\ }\textbf {\bibinfo {volume}
  {44}},\ \bibinfo {pages} {4049} (\bibinfo {year} {2005})}\BibitemShut
  {NoStop}%
\bibitem [{\citenamefont {Zhan}(2009)}]{Zhan:09}%
  \BibitemOpen
  \bibfield  {author} {\bibinfo {author} {\bibfnamefont {Q.}~\bibnamefont
  {Zhan}},\ }\bibfield  {title} {\bibinfo {title} {Cylindrical vector beams:
  from mathematical concepts to applications},\ }\href
  {https://doi.org/10.1364/AOP.1.000001} {\bibfield  {journal} {\bibinfo
  {journal} {Adv. Opt. Photon.}\ }\textbf {\bibinfo {volume} {1}},\ \bibinfo
  {pages} {1} (\bibinfo {year} {2009})}\BibitemShut {NoStop}%
\bibitem [{\citenamefont {Rosales-Guzmán}\ \emph {et~al.}(2018)\citenamefont
  {Rosales-Guzmán}, \citenamefont {Ndagano},\ and\ \citenamefont
  {Forbes}}]{Rosales-Guzmán_2018}%
  \BibitemOpen
  \bibfield  {author} {\bibinfo {author} {\bibfnamefont {C.}~\bibnamefont
  {Rosales-Guzmán}}, \bibinfo {author} {\bibfnamefont {B.}~\bibnamefont
  {Ndagano}},\ and\ \bibinfo {author} {\bibfnamefont {A.}~\bibnamefont
  {Forbes}},\ }\bibfield  {title} {\bibinfo {title} {A review of complex vector
  light fields and their applications},\ }\href
  {https://doi.org/10.1088/2040-8986/aaeb7d} {\bibfield  {journal} {\bibinfo
  {journal} {J. Opt.}\ }\textbf {\bibinfo {volume} {20}},\ \bibinfo {pages}
  {123001} (\bibinfo {year} {2018})}\BibitemShut {NoStop}%
\bibitem [{\citenamefont {Rubano}\ \emph {et~al.}(2019)\citenamefont {Rubano},
  \citenamefont {Cardano}, \citenamefont {Piccirillo},\ and\ \citenamefont
  {Marrucci}}]{Rubano2019}%
  \BibitemOpen
  \bibfield  {author} {\bibinfo {author} {\bibfnamefont {A.}~\bibnamefont
  {Rubano}}, \bibinfo {author} {\bibfnamefont {F.}~\bibnamefont {Cardano}},
  \bibinfo {author} {\bibfnamefont {B.}~\bibnamefont {Piccirillo}},\ and\
  \bibinfo {author} {\bibfnamefont {L.}~\bibnamefont {Marrucci}},\ }\bibfield
  {title} {\bibinfo {title} {{Q-plate technology: a progress review
  [Invited]}},\ }\href {https://doi.org/10.1364/JOSAB.36.000D70} {\bibfield
  {journal} {\bibinfo  {journal} {J. Opt. Soc. Am. B}\ }\textbf {\bibinfo
  {volume} {36}},\ \bibinfo {pages} {D70} (\bibinfo {year} {2019})}\BibitemShut
  {NoStop}%
\bibitem [{\citenamefont {{Di Colandrea}}\ \emph
  {et~al.}(2023{\natexlab{b}})\citenamefont {{Di Colandrea}}, \citenamefont
  {Babazadeh}, \citenamefont {Dauphin}, \citenamefont {Massignan},
  \citenamefont {Marrucci},\ and\ \citenamefont {Cardano}}]{DiColandrea2023}%
  \BibitemOpen
  \bibfield  {author} {\bibinfo {author} {\bibfnamefont {F.}~\bibnamefont {{Di
  Colandrea}}}, \bibinfo {author} {\bibfnamefont {A.}~\bibnamefont
  {Babazadeh}}, \bibinfo {author} {\bibfnamefont {A.}~\bibnamefont {Dauphin}},
  \bibinfo {author} {\bibfnamefont {P.}~\bibnamefont {Massignan}}, \bibinfo
  {author} {\bibfnamefont {L.}~\bibnamefont {Marrucci}},\ and\ \bibinfo
  {author} {\bibfnamefont {F.}~\bibnamefont {Cardano}},\ }\bibfield  {title}
  {\bibinfo {title} {{Ultra-long quantum walks via spin–orbit photonics}},\
  }\href {https://doi.org/10.1364/OPTICA.474542} {\bibfield  {journal}
  {\bibinfo  {journal} {Optica}\ }\textbf {\bibinfo {volume} {10}},\ \bibinfo
  {pages} {324} (\bibinfo {year} {2023}{\natexlab{b}})}\BibitemShut {NoStop}%
\bibitem [{\citenamefont {Simon}\ and\ \citenamefont
  {Mukunda}(1990)}]{SIMON1990165}%
  \BibitemOpen
  \bibfield  {author} {\bibinfo {author} {\bibfnamefont {R.}~\bibnamefont
  {Simon}}\ and\ \bibinfo {author} {\bibfnamefont {N.}~\bibnamefont
  {Mukunda}},\ }\bibfield  {title} {\bibinfo {title} {Minimal three-component
  su(2) gadget for polarization optics},\ }\href
  {https://doi.org/https://doi.org/10.1016/0375-9601(90)90732-4} {\bibfield
  {journal} {\bibinfo  {journal} {Phys. Lett. A}\ }\textbf {\bibinfo {volume}
  {143}},\ \bibinfo {pages} {165} (\bibinfo {year} {1990})}\BibitemShut
  {NoStop}%
\bibitem [{\citenamefont {Sit}\ \emph {et~al.}(2017)\citenamefont {Sit},
  \citenamefont {Giner}, \citenamefont {Karimi},\ and\ \citenamefont
  {Lundeen}}]{Sit_2017}%
  \BibitemOpen
  \bibfield  {author} {\bibinfo {author} {\bibfnamefont {A.}~\bibnamefont
  {Sit}}, \bibinfo {author} {\bibfnamefont {L.}~\bibnamefont {Giner}}, \bibinfo
  {author} {\bibfnamefont {E.}~\bibnamefont {Karimi}},\ and\ \bibinfo {author}
  {\bibfnamefont {J.~S.}\ \bibnamefont {Lundeen}},\ }\bibfield  {title}
  {\bibinfo {title} {General lossless spatial polarization transformations},\
  }\href {https://doi.org/10.1088/2040-8986/aa7f65} {\bibfield  {journal}
  {\bibinfo  {journal} {J. Opt.}\ }\textbf {\bibinfo {volume} {19}},\ \bibinfo
  {pages} {094003} (\bibinfo {year} {2017})}\BibitemShut {NoStop}%
\bibitem [{\citenamefont {Fl{\"{a}}schner}\ \emph {et~al.}(2016)\citenamefont
  {Fl{\"{a}}schner}, \citenamefont {Rem}, \citenamefont {Tarnowski},
  \citenamefont {Vogel}, \citenamefont {L{\"{u}}hmann}, \citenamefont
  {Sengstock},\ and\ \citenamefont {Weitenberg}}]{Flaschner2016}%
  \BibitemOpen
  \bibfield  {author} {\bibinfo {author} {\bibfnamefont {N.}~\bibnamefont
  {Fl{\"{a}}schner}}, \bibinfo {author} {\bibfnamefont {B.~S.}\ \bibnamefont
  {Rem}}, \bibinfo {author} {\bibfnamefont {M.}~\bibnamefont {Tarnowski}},
  \bibinfo {author} {\bibfnamefont {D.}~\bibnamefont {Vogel}}, \bibinfo
  {author} {\bibfnamefont {D.-S.}\ \bibnamefont {L{\"{u}}hmann}}, \bibinfo
  {author} {\bibfnamefont {K.}~\bibnamefont {Sengstock}},\ and\ \bibinfo
  {author} {\bibfnamefont {C.}~\bibnamefont {Weitenberg}},\ }\bibfield  {title}
  {\bibinfo {title} {{Experimental reconstruction of the Berry curvature in a
  Floquet Bloch band}},\ }\href {https://doi.org/10.1126/science.aad4568}
  {\bibfield  {journal} {\bibinfo  {journal} {Science}\ }\textbf {\bibinfo
  {volume} {352}},\ \bibinfo {pages} {1091} (\bibinfo {year}
  {2016})}\BibitemShut {NoStop}%
\bibitem [{\citenamefont {Tarnowski}\ \emph {et~al.}(2019)\citenamefont
  {Tarnowski}, \citenamefont {{\"{U}}nal}, \citenamefont {Fl{\"{a}}schner},
  \citenamefont {Rem}, \citenamefont {Eckardt}, \citenamefont {Sengstock},\
  and\ \citenamefont {Weitenberg}}]{Tarnowski2019}%
  \BibitemOpen
  \bibfield  {author} {\bibinfo {author} {\bibfnamefont {M.}~\bibnamefont
  {Tarnowski}}, \bibinfo {author} {\bibfnamefont {F.~N.}\ \bibnamefont
  {{\"{U}}nal}}, \bibinfo {author} {\bibfnamefont {N.}~\bibnamefont
  {Fl{\"{a}}schner}}, \bibinfo {author} {\bibfnamefont {B.~S.}\ \bibnamefont
  {Rem}}, \bibinfo {author} {\bibfnamefont {A.}~\bibnamefont {Eckardt}},
  \bibinfo {author} {\bibfnamefont {K.}~\bibnamefont {Sengstock}},\ and\
  \bibinfo {author} {\bibfnamefont {C.}~\bibnamefont {Weitenberg}},\ }\bibfield
   {title} {\bibinfo {title} {{Measuring topology from dynamics by obtaining
  the Chern number from a linking number}},\ }\href
  {https://doi.org/10.1038/s41467-019-09668-y} {\bibfield  {journal} {\bibinfo
  {journal} {Nat. Commun.}\ }\textbf {\bibinfo {volume} {10}},\ \bibinfo
  {pages} {1728} (\bibinfo {year} {2019})}\BibitemShut {NoStop}%
\bibitem [{\citenamefont {Yi}\ \emph {et~al.}(2023)\citenamefont {Yi},
  \citenamefont {Yu}, \citenamefont {Yuan}, \citenamefont {Jiao}, \citenamefont
  {Yang}, \citenamefont {Jiang}, \citenamefont {Zhang}, \citenamefont {Chen},\
  and\ \citenamefont {Pan}}]{PhysRevResearch.5.L032016}%
  \BibitemOpen
  \bibfield  {author} {\bibinfo {author} {\bibfnamefont {C.-R.}\ \bibnamefont
  {Yi}}, \bibinfo {author} {\bibfnamefont {J.}~\bibnamefont {Yu}}, \bibinfo
  {author} {\bibfnamefont {H.}~\bibnamefont {Yuan}}, \bibinfo {author}
  {\bibfnamefont {R.-H.}\ \bibnamefont {Jiao}}, \bibinfo {author}
  {\bibfnamefont {Y.-M.}\ \bibnamefont {Yang}}, \bibinfo {author}
  {\bibfnamefont {X.}~\bibnamefont {Jiang}}, \bibinfo {author} {\bibfnamefont
  {J.-Y.}\ \bibnamefont {Zhang}}, \bibinfo {author} {\bibfnamefont
  {S.}~\bibnamefont {Chen}},\ and\ \bibinfo {author} {\bibfnamefont {J.-W.}\
  \bibnamefont {Pan}},\ }\bibfield  {title} {\bibinfo {title} {Extracting the
  quantum geometric tensor of an optical raman lattice by bloch-state
  tomography},\ }\href {https://doi.org/10.1103/PhysRevResearch.5.L032016}
  {\bibfield  {journal} {\bibinfo  {journal} {Phys. Rev. Res.}\ }\textbf
  {\bibinfo {volume} {5}},\ \bibinfo {pages} {L032016} (\bibinfo {year}
  {2023})}\BibitemShut {NoStop}%
\bibitem [{\citenamefont {Ronneberger}\ \emph {et~al.}(2015)\citenamefont
  {Ronneberger}, \citenamefont {Fischer},\ and\ \citenamefont
  {Brox}}]{ronneberger2015u}%
  \BibitemOpen
  \bibfield  {author} {\bibinfo {author} {\bibfnamefont {O.}~\bibnamefont
  {Ronneberger}}, \bibinfo {author} {\bibfnamefont {P.}~\bibnamefont
  {Fischer}},\ and\ \bibinfo {author} {\bibfnamefont {T.}~\bibnamefont
  {Brox}},\ }\href
  {https://link.springer.com/chapter/10.1007/978-3-319-24574-4_28} {\emph
  {\bibinfo {title} {U-Net: Convolutional Networks for Biomedical Image
  Segmentation}}}\ (\bibinfo  {publisher} {Springer International Publishing},\
  \bibinfo {address} {Cham},\ \bibinfo {year} {2015})\ pp.\ \bibinfo {pages}
  {234--241}\BibitemShut {NoStop}%
\bibitem [{\citenamefont {Zhao}\ \emph {et~al.}(2023)\citenamefont {Zhao},
  \citenamefont {Gong}, \citenamefont {Zhang}, \citenamefont {Yao},\ and\
  \citenamefont {Chen}}]{zhao2023physics}%
  \BibitemOpen
  \bibfield  {author} {\bibinfo {author} {\bibfnamefont {X.}~\bibnamefont
  {Zhao}}, \bibinfo {author} {\bibfnamefont {Z.}~\bibnamefont {Gong}}, \bibinfo
  {author} {\bibfnamefont {Y.}~\bibnamefont {Zhang}}, \bibinfo {author}
  {\bibfnamefont {W.}~\bibnamefont {Yao}},\ and\ \bibinfo {author}
  {\bibfnamefont {X.}~\bibnamefont {Chen}},\ }\bibfield  {title} {\bibinfo
  {title} {Physics-informed convolutional neural networks for temperature field
  prediction of heat source layout without labeled data},\ }\href
  {https://www.sciencedirect.com/science/article/abs/pii/S0952197622005061}
  {\bibfield  {journal} {\bibinfo  {journal} {Eng. Appl. Artif. Intell.}\
  }\textbf {\bibinfo {volume} {117}},\ \bibinfo {pages} {105516} (\bibinfo
  {year} {2023})}\BibitemShut {NoStop}%
\bibitem [{\citenamefont {Wang}\ \emph {et~al.}(2024)\citenamefont {Wang},
  \citenamefont {Song}, \citenamefont {Wang}, \citenamefont {Ren},
  \citenamefont {Zhao}, \citenamefont {Dou}, \citenamefont {Di}, \citenamefont
  {Barbastathis}, \citenamefont {Zhou}, \citenamefont {Zhao},\ and\
  \citenamefont {Lam}}]{wang2024use}%
  \BibitemOpen
  \bibfield  {author} {\bibinfo {author} {\bibfnamefont {K.}~\bibnamefont
  {Wang}}, \bibinfo {author} {\bibfnamefont {L.}~\bibnamefont {Song}}, \bibinfo
  {author} {\bibfnamefont {C.}~\bibnamefont {Wang}}, \bibinfo {author}
  {\bibfnamefont {Z.}~\bibnamefont {Ren}}, \bibinfo {author} {\bibfnamefont
  {G.}~\bibnamefont {Zhao}}, \bibinfo {author} {\bibfnamefont {J.}~\bibnamefont
  {Dou}}, \bibinfo {author} {\bibfnamefont {J.}~\bibnamefont {Di}}, \bibinfo
  {author} {\bibfnamefont {G.}~\bibnamefont {Barbastathis}}, \bibinfo {author}
  {\bibfnamefont {R.}~\bibnamefont {Zhou}}, \bibinfo {author} {\bibfnamefont
  {J.}~\bibnamefont {Zhao}},\ and\ \bibinfo {author} {\bibfnamefont {E.~Y.}\
  \bibnamefont {Lam}},\ }\bibfield  {title} {\bibinfo {title} {On the use of
  deep learning for phase recovery},\ }\href
  {https://www.nature.com/articles/s41377-023-01340-x} {\bibfield  {journal}
  {\bibinfo  {journal} {Light Sci. Appl.}\ }\textbf {\bibinfo {volume} {13}},\
  \bibinfo {pages} {4} (\bibinfo {year} {2024})}\BibitemShut {NoStop}%
\bibitem [{\citenamefont {Proppe}\ \emph {et~al.}()\citenamefont {Proppe},
  \citenamefont {Thekkadath}, \citenamefont {England}, \citenamefont {Bustard},
  \citenamefont {Bouchard}, \citenamefont {Lundeen},\ and\ \citenamefont
  {Sussman}}]{proppe20243d}%
  \BibitemOpen
  \bibfield  {author} {\bibinfo {author} {\bibfnamefont {A.~H.}\ \bibnamefont
  {Proppe}}, \bibinfo {author} {\bibfnamefont {G.}~\bibnamefont {Thekkadath}},
  \bibinfo {author} {\bibfnamefont {D.}~\bibnamefont {England}}, \bibinfo
  {author} {\bibfnamefont {P.~J.}\ \bibnamefont {Bustard}}, \bibinfo {author}
  {\bibfnamefont {F.}~\bibnamefont {Bouchard}}, \bibinfo {author}
  {\bibfnamefont {J.~S.}\ \bibnamefont {Lundeen}},\ and\ \bibinfo {author}
  {\bibfnamefont {B.~J.}\ \bibnamefont {Sussman}},\ }\bibfield  {title}
  {\bibinfo {title} {3d-2d neural nets for phase retrieval in noisy
  interferometric imaging},\ }\href {https://arxiv.org/abs/2402.06063} {\
  }\Eprint {https://arxiv.org/abs/arXiv:2402.06063} {arXiv:2402.06063}
  \BibitemShut {NoStop}%
\bibitem [{\citenamefont {Piccirillo}\ \emph {et~al.}(2010)\citenamefont
  {Piccirillo}, \citenamefont {D'Ambrosio}, \citenamefont {Slussarenko},
  \citenamefont {Marrucci},\ and\ \citenamefont {Santamato}}]{Piccirillo2010}%
  \BibitemOpen
  \bibfield  {author} {\bibinfo {author} {\bibfnamefont {B.}~\bibnamefont
  {Piccirillo}}, \bibinfo {author} {\bibfnamefont {V.}~\bibnamefont
  {D'Ambrosio}}, \bibinfo {author} {\bibfnamefont {S.}~\bibnamefont
  {Slussarenko}}, \bibinfo {author} {\bibfnamefont {L.}~\bibnamefont
  {Marrucci}},\ and\ \bibinfo {author} {\bibfnamefont {E.}~\bibnamefont
  {Santamato}},\ }\bibfield  {title} {\bibinfo {title} {{Photon spin-to-orbital
  angular momentum conversion via an electrically tunable q-plate}},\ }\href
  {https://doi.org/10.1063/1.3527083} {\bibfield  {journal} {\bibinfo
  {journal} {Appl. Phys. Lett.}\ }\textbf {\bibinfo {volume} {97}},\ \bibinfo
  {pages} {241104} (\bibinfo {year} {2010})}\BibitemShut {NoStop}%
\bibitem [{\citenamefont {D'Errico}\ \emph {et~al.}(2020)\citenamefont
  {D'Errico}, \citenamefont {Cardano}, \citenamefont {Maffei}, \citenamefont
  {Dauphin}, \citenamefont {Barboza}, \citenamefont {Esposito}, \citenamefont
  {Piccirillo}, \citenamefont {Lewenstein}, \citenamefont {Massignan},\ and\
  \citenamefont {Marrucci}}]{DErrico2020}%
  \BibitemOpen
  \bibfield  {author} {\bibinfo {author} {\bibfnamefont {A.}~\bibnamefont
  {D'Errico}}, \bibinfo {author} {\bibfnamefont {F.}~\bibnamefont {Cardano}},
  \bibinfo {author} {\bibfnamefont {M.}~\bibnamefont {Maffei}}, \bibinfo
  {author} {\bibfnamefont {A.}~\bibnamefont {Dauphin}}, \bibinfo {author}
  {\bibfnamefont {R.}~\bibnamefont {Barboza}}, \bibinfo {author} {\bibfnamefont
  {C.}~\bibnamefont {Esposito}}, \bibinfo {author} {\bibfnamefont
  {B.}~\bibnamefont {Piccirillo}}, \bibinfo {author} {\bibfnamefont
  {M.}~\bibnamefont {Lewenstein}}, \bibinfo {author} {\bibfnamefont
  {P.}~\bibnamefont {Massignan}},\ and\ \bibinfo {author} {\bibfnamefont
  {L.}~\bibnamefont {Marrucci}},\ }\bibfield  {title} {\bibinfo {title}
  {{Two-dimensional topological quantum walks in the momentum space of
  structured light}},\ }\href
  {https://doi.org/https://dx.doi.org/10.1364/OPTICA.365028} {\bibfield
  {journal} {\bibinfo  {journal} {Optica}\ }\textbf {\bibinfo {volume} {7}},\
  \bibinfo {pages} {108} (\bibinfo {year} {2020})}\BibitemShut {NoStop}%
\bibitem [{\citenamefont {Colandrea}\ \emph {et~al.}()\citenamefont
  {Colandrea}, \citenamefont {Dehghan}, \citenamefont {D'Errico},\ and\
  \citenamefont {Karimi}}]{DiColandreaFQPT}%
  \BibitemOpen
  \bibfield  {author} {\bibinfo {author} {\bibfnamefont {F.~D.}\ \bibnamefont
  {Colandrea}}, \bibinfo {author} {\bibfnamefont {N.}~\bibnamefont {Dehghan}},
  \bibinfo {author} {\bibfnamefont {A.}~\bibnamefont {D'Errico}},\ and\
  \bibinfo {author} {\bibfnamefont {E.}~\bibnamefont {Karimi}},\ }\bibfield
  {title} {\bibinfo {title} {Fourier quantum process tomography},\ }\href
  {https://arxiv.org/abs/2312.13458} {\ }\Eprint
  {https://arxiv.org/abs/arXiv:2312.13458} {arXiv:2312.13458} \BibitemShut
  {NoStop}%
\bibitem [{\citenamefont {Marrucci}\ \emph {et~al.}(2006)\citenamefont
  {Marrucci}, \citenamefont {Manzo},\ and\ \citenamefont
  {Paparo}}]{marrucciqplate}%
  \BibitemOpen
  \bibfield  {author} {\bibinfo {author} {\bibfnamefont {L.}~\bibnamefont
  {Marrucci}}, \bibinfo {author} {\bibfnamefont {C.}~\bibnamefont {Manzo}},\
  and\ \bibinfo {author} {\bibfnamefont {D.}~\bibnamefont {Paparo}},\
  }\bibfield  {title} {\bibinfo {title} {Optical spin-to-orbital angular
  momentum conversion in inhomogeneous anisotropic media},\ }\href
  {https://doi.org/10.1103/PhysRevLett.96.163905} {\bibfield  {journal}
  {\bibinfo  {journal} {Phys. Rev. Lett.}\ }\textbf {\bibinfo {volume} {96}},\
  \bibinfo {pages} {163905} (\bibinfo {year} {2006})}\BibitemShut {NoStop}%
\bibitem [{\citenamefont {Wang}\ \emph {et~al.}(2023)\citenamefont {Wang},
  \citenamefont {Fu}, \citenamefont {Mao}, \citenamefont {Qie}, \citenamefont
  {Stone},\ and\ \citenamefont {Yang}}]{Wang:23}%
  \BibitemOpen
  \bibfield  {author} {\bibinfo {author} {\bibfnamefont {C.}~\bibnamefont
  {Wang}}, \bibinfo {author} {\bibfnamefont {Z.}~\bibnamefont {Fu}}, \bibinfo
  {author} {\bibfnamefont {W.}~\bibnamefont {Mao}}, \bibinfo {author}
  {\bibfnamefont {J.}~\bibnamefont {Qie}}, \bibinfo {author} {\bibfnamefont
  {A.~D.}\ \bibnamefont {Stone}},\ and\ \bibinfo {author} {\bibfnamefont
  {L.}~\bibnamefont {Yang}},\ }\bibfield  {title} {\bibinfo {title}
  {Non-hermitian optics and photonics: from classical to quantum},\ }\href
  {https://doi.org/10.1364/AOP.475477} {\bibfield  {journal} {\bibinfo
  {journal} {Adv. Opt. Photon.}\ }\textbf {\bibinfo {volume} {15}},\ \bibinfo
  {pages} {442} (\bibinfo {year} {2023})}\BibitemShut {NoStop}%
\bibitem [{\citenamefont {Zhong}\ \emph {et~al.}(2020)\citenamefont {Zhong},
  \citenamefont {Wang}, \citenamefont {Deng}, \citenamefont {Chen},
  \citenamefont {Peng}, \citenamefont {Luo}, \citenamefont {Qin}, \citenamefont
  {Wu}, \citenamefont {Ding}, \citenamefont {Hu}, \citenamefont {Hu},
  \citenamefont {Yang}, \citenamefont {Zhang}, \citenamefont {Li},
  \citenamefont {Li}, \citenamefont {Jiang}, \citenamefont {Gan}, \citenamefont
  {Yang}, \citenamefont {You}, \citenamefont {Wang}, \citenamefont {Li},
  \citenamefont {Liu}, \citenamefont {Lu},\ and\ \citenamefont
  {Pan}}]{zhong2020quantum}%
  \BibitemOpen
  \bibfield  {author} {\bibinfo {author} {\bibfnamefont {H.-S.}\ \bibnamefont
  {Zhong}}, \bibinfo {author} {\bibfnamefont {H.}~\bibnamefont {Wang}},
  \bibinfo {author} {\bibfnamefont {Y.-H.}\ \bibnamefont {Deng}}, \bibinfo
  {author} {\bibfnamefont {M.-C.}\ \bibnamefont {Chen}}, \bibinfo {author}
  {\bibfnamefont {L.-C.}\ \bibnamefont {Peng}}, \bibinfo {author}
  {\bibfnamefont {Y.-H.}\ \bibnamefont {Luo}}, \bibinfo {author} {\bibfnamefont
  {J.}~\bibnamefont {Qin}}, \bibinfo {author} {\bibfnamefont {D.}~\bibnamefont
  {Wu}}, \bibinfo {author} {\bibfnamefont {X.}~\bibnamefont {Ding}}, \bibinfo
  {author} {\bibfnamefont {Y.}~\bibnamefont {Hu}}, \bibinfo {author}
  {\bibfnamefont {P.}~\bibnamefont {Hu}}, \bibinfo {author} {\bibfnamefont
  {X.-Y.}\ \bibnamefont {Yang}}, \bibinfo {author} {\bibfnamefont {W.-J.}\
  \bibnamefont {Zhang}}, \bibinfo {author} {\bibfnamefont {H.}~\bibnamefont
  {Li}}, \bibinfo {author} {\bibfnamefont {Y.}~\bibnamefont {Li}}, \bibinfo
  {author} {\bibfnamefont {X.}~\bibnamefont {Jiang}}, \bibinfo {author}
  {\bibfnamefont {L.}~\bibnamefont {Gan}}, \bibinfo {author} {\bibfnamefont
  {G.}~\bibnamefont {Yang}}, \bibinfo {author} {\bibfnamefont {L.}~\bibnamefont
  {You}}, \bibinfo {author} {\bibfnamefont {Z.}~\bibnamefont {Wang}}, \bibinfo
  {author} {\bibfnamefont {L.}~\bibnamefont {Li}}, \bibinfo {author}
  {\bibfnamefont {N.-L.}\ \bibnamefont {Liu}}, \bibinfo {author} {\bibfnamefont
  {C.-Y.}\ \bibnamefont {Lu}},\ and\ \bibinfo {author} {\bibfnamefont {J.-W.}\
  \bibnamefont {Pan}},\ }\bibfield  {title} {\bibinfo {title} {Quantum
  computational advantage using photons},\ }\href
  {https://doi.org/10.1126/science.abe8770} {\bibfield  {journal} {\bibinfo
  {journal} {Science}\ }\textbf {\bibinfo {volume} {370}},\ \bibinfo {pages}
  {1460} (\bibinfo {year} {2020})}\BibitemShut {NoStop}%
\bibitem [{\citenamefont {Abadi}\ \emph {et~al.}(2015)\citenamefont {Abadi},
  \citenamefont {Agarwal}, \citenamefont {Barham}, \citenamefont {Brevdo},
  \citenamefont {Chen}, \citenamefont {Citro}, \citenamefont {Corrado},
  \citenamefont {Davis}, \citenamefont {Dean}, \citenamefont {Devin},
  \citenamefont {Ghemawat}, \citenamefont {Goodfellow}, \citenamefont {Harp},
  \citenamefont {Irving}, \citenamefont {Isard}, \citenamefont {Jia},
  \citenamefont {Jozefowicz}, \citenamefont {Kaiser}, \citenamefont {Kudlur},
  \citenamefont {Levenberg}, \citenamefont {Man\'{e}}, \citenamefont {Monga},
  \citenamefont {Moore}, \citenamefont {Murray}, \citenamefont {Olah},
  \citenamefont {Schuster}, \citenamefont {Shlens}, \citenamefont {Steiner},
  \citenamefont {Sutskever}, \citenamefont {Talwar}, \citenamefont {Tucker},
  \citenamefont {Vanhoucke}, \citenamefont {Vasudevan}, \citenamefont
  {Vi\'{e}gas}, \citenamefont {Vinyals}, \citenamefont {Warden}, \citenamefont
  {Wattenberg}, \citenamefont {Wicke}, \citenamefont {Yu},\ and\ \citenamefont
  {Zheng}}]{tensorflow2015-whitepaper``}%
  \BibitemOpen
  \bibfield  {author} {\bibinfo {author} {\bibfnamefont {M.}~\bibnamefont
  {Abadi}}, \bibinfo {author} {\bibfnamefont {A.}~\bibnamefont {Agarwal}},
  \bibinfo {author} {\bibfnamefont {P.}~\bibnamefont {Barham}}, \bibinfo
  {author} {\bibfnamefont {E.}~\bibnamefont {Brevdo}}, \bibinfo {author}
  {\bibfnamefont {Z.}~\bibnamefont {Chen}}, \bibinfo {author} {\bibfnamefont
  {C.}~\bibnamefont {Citro}}, \bibinfo {author} {\bibfnamefont {G.~S.}\
  \bibnamefont {Corrado}}, \bibinfo {author} {\bibfnamefont {A.}~\bibnamefont
  {Davis}}, \bibinfo {author} {\bibfnamefont {J.}~\bibnamefont {Dean}},
  \bibinfo {author} {\bibfnamefont {M.}~\bibnamefont {Devin}}, \bibinfo
  {author} {\bibfnamefont {S.}~\bibnamefont {Ghemawat}}, \bibinfo {author}
  {\bibfnamefont {I.}~\bibnamefont {Goodfellow}}, \bibinfo {author}
  {\bibfnamefont {A.}~\bibnamefont {Harp}}, \bibinfo {author} {\bibfnamefont
  {G.}~\bibnamefont {Irving}}, \bibinfo {author} {\bibfnamefont
  {M.}~\bibnamefont {Isard}}, \bibinfo {author} {\bibfnamefont
  {Y.}~\bibnamefont {Jia}}, \bibinfo {author} {\bibfnamefont {R.}~\bibnamefont
  {Jozefowicz}}, \bibinfo {author} {\bibfnamefont {L.}~\bibnamefont {Kaiser}},
  \bibinfo {author} {\bibfnamefont {M.}~\bibnamefont {Kudlur}}, \bibinfo
  {author} {\bibfnamefont {J.}~\bibnamefont {Levenberg}}, \bibinfo {author}
  {\bibfnamefont {D.}~\bibnamefont {Man\'{e}}}, \bibinfo {author}
  {\bibfnamefont {R.}~\bibnamefont {Monga}}, \bibinfo {author} {\bibfnamefont
  {S.}~\bibnamefont {Moore}}, \bibinfo {author} {\bibfnamefont
  {D.}~\bibnamefont {Murray}}, \bibinfo {author} {\bibfnamefont
  {C.}~\bibnamefont {Olah}}, \bibinfo {author} {\bibfnamefont {M.}~\bibnamefont
  {Schuster}}, \bibinfo {author} {\bibfnamefont {J.}~\bibnamefont {Shlens}},
  \bibinfo {author} {\bibfnamefont {B.}~\bibnamefont {Steiner}}, \bibinfo
  {author} {\bibfnamefont {I.}~\bibnamefont {Sutskever}}, \bibinfo {author}
  {\bibfnamefont {K.}~\bibnamefont {Talwar}}, \bibinfo {author} {\bibfnamefont
  {P.}~\bibnamefont {Tucker}}, \bibinfo {author} {\bibfnamefont
  {V.}~\bibnamefont {Vanhoucke}}, \bibinfo {author} {\bibfnamefont
  {V.}~\bibnamefont {Vasudevan}}, \bibinfo {author} {\bibfnamefont
  {F.}~\bibnamefont {Vi\'{e}gas}}, \bibinfo {author} {\bibfnamefont
  {O.}~\bibnamefont {Vinyals}}, \bibinfo {author} {\bibfnamefont
  {P.}~\bibnamefont {Warden}}, \bibinfo {author} {\bibfnamefont
  {M.}~\bibnamefont {Wattenberg}}, \bibinfo {author} {\bibfnamefont
  {M.}~\bibnamefont {Wicke}}, \bibinfo {author} {\bibfnamefont
  {Y.}~\bibnamefont {Yu}},\ and\ \bibinfo {author} {\bibfnamefont
  {X.}~\bibnamefont {Zheng}},\ }\href {https://www.tensorflow.org/} {\bibinfo
  {title} {{TensorFlow}: Large-scale machine learning on heterogeneous
  systems}} (\bibinfo {year} {2015}),\ \bibinfo {note} {software available from
  tensorflow.org}\BibitemShut {NoStop}%
\bibitem [{\citenamefont {Kingma}\ and\ \citenamefont {Ba}()}]{kingma2017adam}%
  \BibitemOpen
  \bibfield  {author} {\bibinfo {author} {\bibfnamefont {D.~P.}\ \bibnamefont
  {Kingma}}\ and\ \bibinfo {author} {\bibfnamefont {J.}~\bibnamefont {Ba}},\
  }\bibfield  {title} {\bibinfo {title} {Adam: A method for stochastic
  optimization},\ }\href {https://arxiv.org/abs/1412.6980} {\ }\Eprint
  {https://arxiv.org/abs/arXiv:1412.6980} {arXiv:1412.6980} \BibitemShut
  {NoStop}%
\bibitem [{\citenamefont {Taylor}\ \emph {et~al.}()\citenamefont {Taylor},
  \citenamefont {Kardas}, \citenamefont {Cucurull}, \citenamefont {Scialom},
  \citenamefont {Hartshorn}, \citenamefont {Saravia}, \citenamefont {Poulton},
  \citenamefont {Kerkez},\ and\ \citenamefont {Stojnic}}]{taylor2022galactica}%
  \BibitemOpen
  \bibfield  {author} {\bibinfo {author} {\bibfnamefont {R.}~\bibnamefont
  {Taylor}}, \bibinfo {author} {\bibfnamefont {M.}~\bibnamefont {Kardas}},
  \bibinfo {author} {\bibfnamefont {G.}~\bibnamefont {Cucurull}}, \bibinfo
  {author} {\bibfnamefont {T.}~\bibnamefont {Scialom}}, \bibinfo {author}
  {\bibfnamefont {A.}~\bibnamefont {Hartshorn}}, \bibinfo {author}
  {\bibfnamefont {E.}~\bibnamefont {Saravia}}, \bibinfo {author} {\bibfnamefont
  {A.}~\bibnamefont {Poulton}}, \bibinfo {author} {\bibfnamefont
  {V.}~\bibnamefont {Kerkez}},\ and\ \bibinfo {author} {\bibfnamefont
  {R.}~\bibnamefont {Stojnic}},\ }\bibfield  {title} {\bibinfo {title}
  {Galactica: A large language model for science},\ }\href
  {https://arxiv.org/abs/2211.09085} {\ }\Eprint
  {https://arxiv.org/abs/arXiv:2211.09085} {arXiv:2211.09085} \BibitemShut
  {NoStop}%
\bibitem [{\citenamefont {Weiss}\ \emph {et~al.}(2016)\citenamefont {Weiss},
  \citenamefont {Khoshgoftaar},\ and\ \citenamefont {Wang}}]{TransferLearning}%
  \BibitemOpen
  \bibfield  {author} {\bibinfo {author} {\bibfnamefont {K.}~\bibnamefont
  {Weiss}}, \bibinfo {author} {\bibfnamefont {T.~M.}\ \bibnamefont
  {Khoshgoftaar}},\ and\ \bibinfo {author} {\bibfnamefont {D.}~\bibnamefont
  {Wang}},\ }\bibfield  {title} {\bibinfo {title} {A survey of transfer
  learning},\ }\href
  {https://journalofbigdata.springeropen.com/articles/10.1186/s40537-016-0043-6}
  {\bibfield  {journal} {\bibinfo  {journal} {J. Big Data}\ }\textbf {\bibinfo
  {volume} {3}},\ \bibinfo {pages} {1} (\bibinfo {year} {2016})}\BibitemShut
  {NoStop}%
\end{thebibliography}%

\end{document}